\pdfoutput=1
\RequirePackage{etoolbox}

\documentclass[noinfoline]{imsart}

\usepackage{amsthm}
\usepackage{amsmath}
\usepackage{amsfonts}
\usepackage{natbib}
\usepackage[colorlinks,citecolor=blue,urlcolor=blue,filecolor=blue,backref=page]{hyperref}
\usepackage{graphicx}

\usepackage[overload]{abraces}
\usepackage{booktabs}
\usepackage{hyperref}
\usepackage[dvipsnames]{xcolor}

\usepackage{subfig}

\renewcommand{\Pr}{\text{Pr}}
\newcommand{\boldc}{\boldsymbol{c}}

\usepackage{tikz}
\usetikzlibrary{fit,positioning,calc} 

\usepackage{algorithm}
\usepackage{algorithmic}

\makeatletter
\newenvironment{breakablealgorithm}
  {
   \begin{center}
     \refstepcounter{algorithm}
     \hrule height.8pt depth0pt \kern2pt
     \renewcommand{\caption}[2][\relax]{
       {\raggedright\textbf{\ALG@name~\thealgorithm} ##2\par}%
       \ifx\relax##1\relax 
         \addcontentsline{loa}{algorithm}{\protect\numberline{\thealgorithm}##2}%
       \else 
         \addcontentsline{loa}{algorithm}{\protect\numberline{\thealgorithm}##1}%
       \fi
       \kern2pt\hrule\kern2pt
     }
  }{
     \kern2pt\hrule\relax
   \end{center}
  }
\makeatother

\startlocaldefs
\endlocaldefs

\begin{document}


\begin{frontmatter}
 \title{Centered Partition Processes: Informative Priors for Clustering}

 \begin{aug}
  \author{\fnms{Sally} \snm{Paganin}\thanksref{a1}\ead[label=e1]{paganin@stat.unipd.it}},
  \author{\fnms{Amy H.} \snm{Herring}\thanksref{a2}},
  \author{\fnms{Andrew F.} \snm{Olshan}\thanksref{a3}},
  \author{\fnms{David B.} \snm{Dunson}\thanksref{a2}}
  \and
  \author{\fnms{The National Birth Defects Prevention Study}}

  \address{Department of Statistical Sciences, University of Padova, Italy\thanksmark{a1}\\ 
  \printead{e1}}
  \address{ Department of Statistical Science, Duke University, USA\thanksmark{a2}}
  \address{Department of Epidemiology, The University of North Carolina at Chapel Hill, USA\thanksmark{a3}}
  \runauthor{S. Paganin et al.}
  \runtitle{Centered Partition Processes}
 \end{aug}

 \begin{abstract}
  There is a very rich literature proposing Bayesian approaches for clustering starting with a prior probability distribution on partitions.
  Most approaches assume exchangeability, leading to simple representations in terms of Exchangeable Partition Probability Functions (EPPF).
  Gibbs-type priors encompass a broad class of such cases, including Dirichlet and Pitman-Yor processes.
  Even though there have been some proposals to relax the exchangeability assumption, allowing covariate-dependence and partial exchangeability, limited consideration has been given on how to include concrete prior knowledge on the partition.
  For example, we are motivated by an epidemiological application, in which we wish to cluster birth defects into groups and we have prior knowledge of an initial clustering provided by experts.
  As a general approach for including such prior knowledge, we propose a Centered Partition (CP) process that modifies the EPPF to favor partitions close to an initial one.
  Some properties of the CP prior are described, a general algorithm for posterior computation is developed, and we illustrate the methodology through simulation examples and an application to the motivating epidemiology study of birth defects.
 \end{abstract}


\end{frontmatter}


\section{Introduction}\label{sec1_intro}

Clustering is one of the canonical data analysis goals in statistics. There are two main strategies that have been used for clustering; namely, distance and model-based clustering.
Distance-based methods leverage upon a distance metric between data points and do not in general require a generative probability model of the data, while model-based methods rely on discrete mixture models, which model the data in different clusters as arising from kernels having different parameter values. The majority of the model-based literature leans on maximum likelihood estimation, commonly relying on the EM algorithm. Bayesian approaches that aim to approximate a full posterior distribution on the clusters have advantages in terms of uncertainty quantification, while also having the ability to incorporate prior information.

Although this article is motivated by providing a broad new class of methods for improving clustering performance in practice, we were initially motivated by a particular application involving birth defects epidemiology. In this context, there are $N$ different birth defects, which we can index using $i \in \{1,\ldots,N\}$, and we are interested in clustering these birth defects into mechanistic groups. This may be useful, for example, in that birth defects in the same group may have similar coefficients in logistic regression analysis relating different exposures to risk of developing the defect. Investigators have provided us with an initial partition $\boldc_0$ of the defects $\{1,\ldots,N\}$ into groups. It is appealing to combine this prior knowledge with information in the data from a grouped logistic regression to produce a posterior distribution on clusters, which characterizes uncertainty. The motivating question of this article is how to do this, with the resulting method ideally having broader impact to other types of {\em centering} of priors for clustering; for example, we may want to center the prior based on information on the number of clusters or cluster sizes.

With these goals in mind, we start by reviewing the relevant literature on clustering priors. Most of these methods assume {\em exchangeability}, which means that the prior probability of a partition $\boldc$ of $\{1,\ldots,N\}$ into clusters depends only on the number of clusters and the cluster sizes; the indices on the clusters play no role. Under the exchangeability assumption, one can define what is referred to in the literature as an Exchangeable Partition Probability Function (EPPF) \citep{pitman1995}. This EPPF provides a prior distribution on the random partition $\boldc$. One direction to obtain a specific form for the EPPF is to start with a nonparametric Bayesian discrete mixture model with a prior for the mixing measure $P$, and then marginalize over this prior to obtain an induced prior on partitions. Standard choices for $P$, such as the Dirichlet \citep{ferguson1973} and Pitman-Yor process \citep{pitman1997}, lead to relatively simple analytic forms for the EPPF. There has been some recent literature studying extensions to broad classes of Gibbs-type processes \citep{gnedin2006, deblasi2015}, mostly focused on improving flexibility while maintaining the ability to predict the number of new clusters in a future sample of data.

There is also a rich literature on relaxing exchangeability in various ways. Most of the emphasis has been on the case in which a vector of features $\mathbf{x}_i$ is available for index $i$, motivating feature-dependent random partitions models.
Building on the stick-breaking representation of the DP \citep{sethuraman1994}, \cite{maceachern1999, maceachern2000} proposed a class of fixed weight dependent DP (DDP) priors. Applications of this DDP framework have been employed in ANOVA modeling \citep{deiorio2004}, spatial data analysis \citep{gelfand2005}, time series \citep{caron2006} and functional data analysis applications \citep{petrone, scarpa2009} among many others, with some theoretical properties highlighted in \cite{barrientos2012}.

However such fixed weight DDPs lack flexibility in feature-dependent clustering, as noted in \cite{maceachern2000}. This has motivated alternative formulations which allow the mixing weights to change with the features, with some examples including the order-based dependent Dirichlet process \citep{griffin2006}, kernel- \citep{dunson2008}, and probit- \citep{rodriguez2011} stick breaking processes.

Alternative approaches build on random partition models (RPMs), working directly with the probability distribution $p(\boldc)$ on the partition $\boldc$ of indices $\{1,\ldots,N\}$ into clusters. Particular attention has been given to the class of product partition models (PPMs) \citep{barry1992, hartigan1990} in which $p(\boldc)$ can be factorized into a product of cluster-dependent functions, known as \emph{cohesion functions}. A common strategy modifies the cohesion to allow feature-dependence; refer, for examples, to \cite{park2010}, \cite{muller2011}, \cite{blei2011} and \cite{dahl2016}.

Our focus is fundamentally different. In particular, we do not have features $\mathbf{x}_i$ on indices $i$ but have access to an informed prior guess $\boldc_0$ for the partition $\boldc$; other than this information it is plausible to rely on exchangeable priors. To address this problem, we propose a general strategy to modify a baseline EPPF to include centering on $\boldc_0$. In particular, our proposed Centered Partition (CP) process defines the partition prior as proportional to an EPPF multiplied by an exponential factor which depends on a distance function $d(\boldc,\boldc_0)$ measuring how far $\boldc$ is from $\boldc_0$. The proposed framework should be broadly useful in including extra information into EPPFs, which tend to face issues in lacking incorporation of real prior information from applications.

The paper is organized as follows. Section~\ref{sec2:preliminaries} introduces concepts and notation related to Bayesian nonparametric clustering. In Section~\ref{sec3:Centered_pp} we illustrate the general CP process formulation and describe an approach to posterior computation relying on Markov chain Monte Carlo (MCMC). Section~\ref{sec:prior_calibration} proposes a general strategy for prior calibration building on a targeted Monte Carlo procedure. Simulation studies and application to the motivating birth defects epidemiology study are provided in Section~\ref{sec5:nbdps}, with technical details included in an Appendix.

\section{Clustering and Bayesian models}\label{sec2:preliminaries}

This section introduces some concepts related to the representation of the clustering space from a combinatorial perspective, which will be useful to define the Centered Partition process, along with an introduction to Bayesian nonparametric clustering models.

\subsection{Set Partitions}\label{sec2.1}

Let $\boldsymbol{c}$ be a generic clustering of indices $[N] = \{1, \ldots, N\}$. It can be either represented as a vector of indices $\{c_1, \ldots, c_N\}$, with $c_i \in \{1, \ldots, K\}$ for $i = 1,\ldots, N$ and  $c_i = c_j$ when $i$ and $j$ belong to the same cluster, or as a collection of disjoint subsets (blocks) $\{B_1, B_2, \ldots, B_{K}\}$ where $B_k$ contains all the indices of data points in the $k$-th cluster and $K$ is the number of clusters in the sample of size $N$.
From a mathematical perspective $\boldsymbol{c} = \{B_1, \ldots, B_{K}\}$ is a combinatorial object known as \textit{set partition} of $[N]$.
The collection of all possible set partitions of $[N]$, denoted with $\Pi_N$, is known as the \textit{partition lattice}. We refer to \cite{stanley1997} and \cite{davey2002} for an introduction to lattice theory, and to \cite{meila2007} and \cite{wade2018} for a review of the concepts from a more statistical perspective.

According to Knuth in \cite{wilson2013combinatorics}, set partitions seem to have been studied first in Japan around A.D. 1500, due to a popular game in the upper class society known as \textit{genji-ko}; five unknown incense sticks were burned and players were asked to identify which of the scents were the same, and which were different.
Soon diagrams were developed to model all the $52$ outcomes, which corresponds to all the possible set partitions of $N = 5$ elements. First results focused on enumerating the elements of the space. For example, for a fixed number of blocks $K$, the number of ways to assign $N$ elements to $K$ groups is described by the \emph{Stirling number of the second kind}
\begin{equation*}
 \mathcal{S}_{N, K} = \frac{1}{K!} \sum_{j = 0}^K (-1)^j {{K}\choose{j}} (K - j)^N,
\end{equation*}
while the \emph{Bell number} $\mathcal{B}_N = \sum_{K = 1}^N \mathcal{S}_{N, K}$ describes the number of all possible set partitions of $N$ elements.

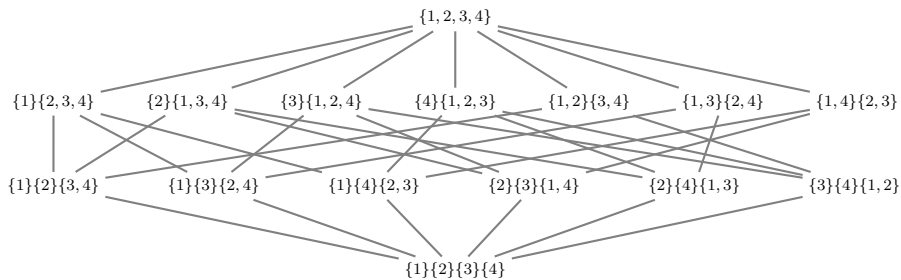
\begin{figure}[!h]
 \centering
 \begin{tikzpicture}[align = center, scale = 0.75, transform shape]
  \node (1) at (0,0) {$\{1,2,3,4\}$};
  \node [below = of 1] (2) {$\{4\}\{1,2,3\}$};
  \node [left=0.7cm of 2] (3)  {$\{3\}\{1,2,4\}$};
  \node [left=0.7cm of 3] (4)  {$\{2\}\{1,3,4\}$};
  \node [left=0.7cm of 4] (5)  {$\{1\}\{2,3,4\}$};

  \node [right =0.7cm of 2] (6) {$\{1,2\}\{3,4\}$};
  \node [right =0.7cm of 6] (7) {$\{1,3\}\{2,4\}$};
  \node [right =0.7cm of 7] (8) {$\{1,4\}\{2,3\}$};

  \node [below = of 5] (9) {$\{1\}\{2\}\{3,4\}$};
  \node [right =1cm of 9] (10) {$\{1\}\{3\}\{2,4\}$};
  \node [right =1cm of 10] (11) {$\{1\}\{4\}\{2,3\}$};
  \node [right =1cm of 11] (12) {$\{2\}\{3\}\{1,4\}$};
  \node [right =1cm of 12] (13) {$\{2\}\{4\}\{1,3\}$};
  \node [right =1cm of 13] (14) {$\{3\}\{4\}\{1,2\}$};
  \node [below =2.5cm of 2] (0) {$\{1\}\{2\}\{3\}\{4\}$};

  \draw [gray,  thick] (1) -- (2) (1) -- (3) (1) -- (4) (1) -- (5) (1) -- (6) (1) -- (7) (1) -- (8);
  \draw [gray,  thick] (5) -- (9) (5) -- (10) (5) -- (11);
  \draw [gray,  thick] (4) -- (9) (4) -- (12) (4) -- (13);
  \draw [gray,  thick] (3) -- (10) (3) -- (12) (3) -- (14);
  \draw [gray,  thick] (2) -- (11) (2) -- (13) (2) -- (14);
  \draw [gray,  thick] (6) -- (9) (6) -- (14);
  \draw [gray,  thick] (7) -- (10) (7) -- (13);
  \draw [gray,  thick] (8) -- (11) (8) -- (12);

  \draw [gray,  thick] (9) -- (0) (10) -- (0) (11) -- (0) (12) -- (0) (13) -- (0) (14) -- (0);

 \end{tikzpicture}
 \caption{Hasse diagram for the lattice of set partitions of $4$ elements. A line is drawn when two partitions have a covering relation. For example  $\{1\}\{2,3,4\}$ is connected with $3$ partitions obtained by splitting the block $\{2,3,4\}$ in every possible way, and with partition $\mathbf{1}$, obtained by merging the two clusters.}\label{fig_hasse4}
\end{figure}

Interest progressively shifted towards characterizing the structure of the space of partitions using the notion of partial order.
Consider $\Pi_N$ endowed with the set containment relation $\leq$, meaning that for $\boldsymbol{c}= \{B_1, \ldots, B_K\}, \boldsymbol{c}^\prime= \{B^\prime_1, \ldots, B^\prime_{K'}\}$ belonging to $\Pi_N$, $\boldsymbol{c} \leq \boldsymbol{c}^\prime$ if for all $i = 1,\ldots, K, B_i \subseteq B^\prime_j$ for some $j \in \{1, \ldots, K^\prime\}$.
Then the space $(\Pi_N, \leq)$ is a \textit{partially ordered set} (poset), which satisfies the following properties:
\begin{enumerate}
 \item Reflexivity: for every $\boldc, \boldc^\prime \in \Pi_N$, $\boldc \leq \boldc$,
 \item Antisymmetry: if $\boldc \leq \boldc^\prime$ and $\boldc^\prime \leq \boldc$ then $\boldc = \boldc^\prime$,
 \item Transitivity: if $\boldc \leq \boldc^\prime$ and $\boldc' \leq \boldc''$, then $\boldc \leq \boldc''$.
\end{enumerate}
Moreover, for any $\boldc, \boldc^\prime \in \Pi_N$, it is said that $\boldc$ is \emph{covered} (or refined) by $\boldc^\prime$ if $\boldc \leq \boldc'$ and there is no $\boldc''$ such that $\boldc < \boldc'' < \boldc^\prime$. Such a relation is indicated by $\boldc \prec \boldc^\prime$.
This covering relation allows one to represent the space of partitions using a \emph{Hasse diagram}, in which the elements of $\Pi_N$ correspond to nodes in a graph and a line is drawn from $\boldc$ to $\boldc^\prime$ when $\boldc \prec \boldc^\prime$; there is a connection from a partition $\boldc$ to another one when the second can be obtained by splitting or merging one of the blocks in $\boldc$. See Fig.~\ref{fig_hasse4} for an example of the Hasse diagram of $\Pi_4$.
Conventionally, the partition with just one cluster is represented at the top of the diagram and denoted as $\boldsymbol{1}$, while the partition having every observation in its own cluster is at the bottom and indicated with $\boldsymbol{0}$.

This representation of the set partitions space $\Pi_N$ as a partially ordered set provides a useful framework to characterize its elements.
As already mentioned, two partitions connected in the Hasse diagram can be obtained from one another by means of a single operation of split or merge; a sequence of connections is a \textit{path}, linking the two extreme partitions $\boldsymbol{0}$ and $\boldsymbol{1}$. A path starting from $\boldsymbol{0}$ connects partitions with an increasing number of blocks, from $1$ to $N$, which is referred as the \emph{rank} of the partition.
Set partitions with the same rank may differ in terms of their \textit{configuration} $\boldsymbol{\Lambda}(\boldc)$, the sequence of block cardinalities $\{|B_1|, \ldots, |B_{K}|\}$, which corresponds to another combinatorial object known as an \textit{integer partition} of $N$. In combinatorics, an integer partition is defined as the multiset of positive integers $\{\lambda_1 \ldots \lambda_K\}$, listed in decreasing order by convention, such that $\sum_{i = 1}^K \lambda_i = N$. Also the associated space of all possible integer partitions $I_N$ is a partially ordered set, making the definition of configuration a poset mapping $\boldsymbol{\Lambda}(\cdot): \boldc \in \Pi_N \rightarrow \boldsymbol{\lambda} \in I_N$.

Finally, the space $\Pi_N$  is a \emph{lattice}, based on the fact that every pair of elements has a \emph{greatest lower bound} (g.l.b.) and a \emph{least upper bound} (l.u.b.) indicated with the ``meet'' $\land$  and the ``join'' $\lor$ operators, i.e. $\boldc \land \boldc' = \text{g.l.b.}(\boldc, \boldc')$ and $\boldc \lor \boldc' = \text{l.u.b.}(\boldc, \boldc')$ and equality holds under a permutation of the cluster labels. An element $\boldc \in \Pi_N$ is an upper bound for a subset $\boldsymbol{S} \subseteq \Pi_N$ if $\boldsymbol{s} \leq \boldc$ for all $\boldsymbol{s} \in \boldsymbol{S}$, and it is the least upper bound for a subset $\boldsymbol{S} \subseteq \Pi_N$ if $\boldc$ is an upper bound for $\boldsymbol{S}$ and $\boldc \leq \boldc^\prime$ for all upper bounds $\boldc^\prime$ of $\boldsymbol{S}$.
The lower bound and the greatest lower bound are defined similarly, and the definition applies also to the elements of the space $I_N$.
Consider, as an example, $\boldc = \{1\}\{2,3,4\}$ and $\boldc' = \{3\}\{1,2,4\}$; their greatest lower bound is $\boldc \land \boldc' =\{1\}\{3\}\{2,4\}$ while the lowest upper bound is $\boldc \lor \boldc' = \{1,2,3,4\}$.
Considering the Hasse diagram in Fig~\ref{fig_hasse4} the g.l.b. and l.u.b. are the two partitions which reach both $\boldc$ and $\boldc'$ through the shortest path, respectively from below and from above.

\subsection{Bayesian mixture models}\label{sec2.2:bnp}
From a statistical perspective, set partitions are key elements in a Bayesian mixture model framework. The main underlying assumption is that observations $y_1, \ldots, y_N$ are independent conditional on the partition $\boldc$, and their joint probability density can be expressed as
\begin{equation}\label{eq:mixture_model}
 p(\mathbf{y} | \boldc, \boldsymbol{\theta}) = \prod_{k =1}^{K} \prod_{i \in B_k} p(y_i | \theta_k ) = \prod_{k =1}^{K} p(\mathbf{y}_k | \theta_k ),
\end{equation}
with $\boldsymbol{\theta} = (\theta_1, \ldots, \theta_K)$ a vector of unknown parameters indexing the distribution of observations $\mathbf{y}_k = \{y_i\}_{i \in B_k}$ for each cluster $k = 1, \ldots, K$.
In a Bayesian formulation, a prior distribution is assigned to each possible partition $\boldc$, leading to a posterior of the form
\begin{equation}
 p(\boldc| \mathbf{y}, \boldsymbol{\theta}) \propto p(\boldc) \prod_{k =1}^{K} p(\mathbf{y}_k | \theta_k ).
\end{equation}
Hence the set partition $\boldc$ is conceived as a random object and elicitation of its prior distribution is a critical issue in Bayesian modeling.

The first distribution one may use, in the absence of prior information, is the uniform distribution, which gives the same probability to every partition with $p(\boldc) = 1/\mathcal{B}_N$; however, even for small values of $N$ the Bell number $\mathcal{B}_N$ is very large, making computation of the posterior intractable even for simple choices of the likelihood. This motivated the definition of alternative prior distributions based on different concepts of uniformity, with the \cite{jensen2008} prior favoring uniform placement of new observations in one of the existing clusters, and \cite{casella2014} proposing a hierarchical uniform prior, which gives equal probability to set partitions having the same configuration.

Usual Bayesian nonparametric procedures build instead on discrete nonparametric priors, i.e. priors that have discrete realization almost surely.
Dirichlet and Pitman-Yor processes are well known to have this property, as does the broader class of Gibbs-type priors.
Any discrete random probability measure $\tilde{p}$ can induce an exchangeable random partition.
Due to the discreteness of the process, $\tilde{p}$ induces a partition of the observations $y_1, \ldots, y_N$ which can be characterized via an Exchangeable Probability Partition Function. For both Dirichlet and Pitman-Yor processes, the EPPF is available in closed form as reported in Table~\ref{tab:eppf} along with the case of the finite mixture model with $\kappa$ components and a symmetric Dirichlet prior with parameters $(\gamma/\kappa, \ldots, \gamma/\kappa)$.
Notice that $\lambda_j = |B_j|$ is the cardinality of the clusters composing the partition, while notation $(x)_r$ is for the rising factorial $x(x + 1)\cdots(x + r -1)$.

\begin{table}
 \centering
 $\begin{array}{ ccc }
   \toprule
   \text{Random probability measure} & \text{Parameters}                                                                                                 & p(\boldc) = \\
   \midrule
   \text{Dirichlet process}          & (\alpha)
                                     & \frac{\alpha^{K}}{(\alpha)_N} \prod_{j = 1}^{K} (\lambda_j - 1)!                                                                \\
   \text{Pitman-Yor process}                 & (\alpha, \sigma)
                                     & \frac{\prod_{j = 1}^{K-1}(\alpha +j\sigma)}{(\alpha +1)_{(N-1)}} \prod_{j = 1}^{K} (1 - \sigma)_{(\lambda_j - 1)}               \\
   \text{Symmetric Dirichlet}        & (\kappa, \gamma)
                                     & \frac{\kappa!}{(\kappa - K)!} \prod_{j = 1}^{K} \frac{\Gamma(\gamma/\kappa + \lambda_j)}{\Gamma(\gamma/\kappa)}                 \\
   \bottomrule
  \end{array}$
 \caption{Exchangeable Partition Probability Function for Dirichlet, Pitman-Yor processes and Symmetric Dirichlet distribution; $\lambda_j = |B_j|$ is the cardinality of the clusters composing the partition, while $(x)_r = x(x + 1)\cdots(x + r -1)$ denotes the rising factorial.}
 \label{tab:eppf}
\end{table}

There is a strong connection with the exchangeable random partitions induced by Gibbs-type priors and product partition models. A product partition model assumes that the prior probability for the partition $\boldc$ has the following form
\begin{equation}
 p(\boldc = \{B_1, \ldots, B_K\}) \propto \prod_{j = 1}^{K} \rho(B_j),
\end{equation}
with $\rho(\cdot)$ known as the cohesion function. The underlying assumption is that the prior distribution for the set partition $\boldc$ can be factorized as the product of functions that depend only on the blocks composing it. Such a definition, in conjunction with formulation \eqref{eq:mixture_model} for the data likelihood, guarantees the property that the posterior distribution for $\boldc$ is still in the class of product partition models.

Distributions in Table~\ref{tab:eppf} are all characterized by a cohesion function that depends on the blocks through their cardinality.
Although the parameters can control the expected number of clusters, this assumption is too strict in many applied contexts in which prior information is available about the grouping. In particular, the same probability is given to partitions with the same configuration but having a totally different composition.

\section{Centered Partition Processes}\label{sec3:Centered_pp}

Our focus is on incorporating structured knowledge about clustering of the finite set of indices $[N] = \{1, \ldots,N\}$ in the prior distribution within a Bayesian mixture model framework.
We consider as a first source of information a given potential clustering, but our approach can also accommodate prior information on summary statistics such as the number of clusters and cluster sizes.

\subsection{General formulation}\label{sec3.1:general}

Assume that a potential clustering $\boldc_0$ is given and we wish to include this information in the prior distribution.
To address this problem, we propose a general strategy to modify a baseline EPPF to shrink towards $\boldc_0$.
In particular, our proposed CP process defines the prior on set partitions as proportional to a baseline EPPF multiplied by a penalization term of the type
\begin{equation}\label{eq_cenex}
 p(\boldc| \boldc_0, \psi) \propto p_0(\boldc) e^{-\psi d(\boldc, \boldc_0)},
\end{equation}
with $\psi > 0$ a penalization parameter, $d(\boldc, \boldc_0)$ a suitable distance measuring how far $\boldc$ is from $\boldc_0$ and $p_0(\boldc)$ indicates a baseline EPPF, that may depend on some parameters that are not of interest at the moment. For $\psi \rightarrow 0$, $p(\boldc| \boldc_0, \psi)$ corresponds to the baseline EPPF $p(\boldc_0)$, while for $\psi \rightarrow \infty$, $p(\boldc = \boldc_0) \rightarrow 1$.

Note that $d(\boldc, \boldc_0)$ takes a finite number of discrete values $\Delta = \{\delta_0, \ldots, \delta_L\}$, with $L$ depending on $\boldc_0$ and on the distance $d(\cdot, \cdot)$. We can define sets of partitions having the same fixed distance from $\boldc_0$ as
\begin{equation}\label{eq:set_c0}
 s_l(\boldc_0) = \{\boldc \in \Pi_N: d(\boldc, \boldc_0) = \delta_l\}, \quad l = 0, 1, \ldots, L.
\end{equation}
Hence, for $\delta_0 = 0$,  $s_0(\boldc_0)$ denotes the set of partitions equal to the base one, meaning that they differ from $\boldc_0$ only by a permutation of the cluster labels.
Then $s_1(\boldc_0)$ denotes the set of partitions with minimum distance $\delta_1$ from $\boldc_0$, $s_2(\boldc_0)$ the set of partitions with the second minimum distance $\delta_2$ from $\boldc_0$ and so on.
The introduced exponential term penalizes equally partitions in the same set $s_l(\boldc_0)$ for a given $\delta_l$, but the resulting probabilities may differ depending on the chosen baseline EPPF.

\subsection{Choices of distance function}\label{sec3:distance}

The proposed CP process modifies a baseline EPPF to include a distance-based penalization term, which aims to shrink the prior distribution towards a prior partition guess. The choice of distance plays a key role in determining the behavior of the prior distribution.
A variety of different distances and indices have been employed in clustering procedures and comparisons. We consider in this paper the Variation of Information (VI), obtained axiomatically in \cite{meila2007} using information theory, and shown to nicely characterize neighborhoods of a given partition by \cite{wade2018}. The Variation of Information is based on the Shannon entropy $H(\cdot)$, and can be computed as
\begin{align*}
 \text{VI}(\boldc, \boldc') & = -H(\boldc) - H(\boldc_0) + 2H(\boldc \land \boldc_0)                       \\
                            & = \sum_{j = 1}^K \frac{\lambda_j}{N} \log \left( \frac{\lambda_j}{N} \right)
 + \sum_{l = 1}^{K'}\frac{\lambda^\prime_l}{N} \log \left(\frac{\lambda^\prime_l}{N}\right)
 - 2 \sum_{j = 1}^K \sum_{l = 1}^{K'} \frac{\lambda^{\land}_{jl}}{N} \log \left( \frac{\lambda^{\land}_{jl}}{N}\right),
\end{align*}
where $\log$ denotes $\log$ base 2, and $\lambda^{\land}_{jl}$ the size of blocks of the intersection $\boldc \land \boldc^\prime$ and hence the number of indices in block $j$ under partition $\boldc$ and block $l$ under $\boldc^\prime$. Notice that VI ranges from $0$ to $\log_2(N)$. Although normalized versions have been proposed \citep{vinh2010}, some desirable properties are lost under normalization.
We refer to \cite{meila2007} and \cite{wade2018} for additional properties and empirical evaluations.

An alternative definition of the VI can be derived from lattice theory, exploiting the concepts provided in Section~\ref{sec2.1}. We refer to \cite{monjardet1981} for general theory about metrics on lattices and ordered sets, and \cite{rossi2015} for a more recent review focused on set partitions.
In general, a distance between two different partitions $\boldc, \boldc^\prime \in \Pi_N$ can be defined by means of the Hasse diagram via the minimum weighted path, which corresponds to the shortest path length when edges are equally weighted. Instead, when edges depend on the entropy function through $w(\boldc, \boldc^\prime) = |H(\boldc) - H(\boldc^\prime)|$, the minimum weighted path between two partitions is the Variation of Information. Notice that two partitions are connected when in a covering relation, then $\boldc \land \boldc^\prime$ is either equal to $\boldc$ or $\boldc^\prime$ and $VI(\boldc, \boldc^\prime) = w(\boldc, \boldc^\prime)$.
The minimum weight $w(\boldc, \boldc^\prime)$ corresponds to $2/N$ which is attained when two singleton clusters are merged, or conversely, a cluster consisting of two points is split \citep[see][]{meila2007}.

\subsection{Effect of the prior penalization}\label{sec:3.3_prior}

We first consider the important special case in which the baseline EPPF is $p_0(\boldc) = 1/\mathcal{B}_N$ and the CP process reduces to $p(\boldc| \boldc_0, \psi) \propto \exp\{-\psi d(\boldc, \boldc_0)\}$ with equation~\eqref{eq_cenex} simplifying to
\begin{equation}\label{eq:analitic_cp}
 p(\boldc| \boldc_0, \psi) =  \frac{ e^{-\psi \delta_l}}{\sum_{u = 0}^L  |s_u(\boldc_0)| e^{-\psi \delta_u} },  \quad \text{for } \boldc \in s_l(\boldc_0),  \quad l = 0, 1, \ldots, L,
\end{equation}
where $| \cdot |$ indicates the cardinality and $s_l(\boldc_0)$ is defined in \eqref{eq:set_c0}.

\begin{figure}[!ht]
 \subfloat[$\boldc_0 = \{1,2,3,4,5\}$]{\includegraphics[width = 0.48\textwidth]{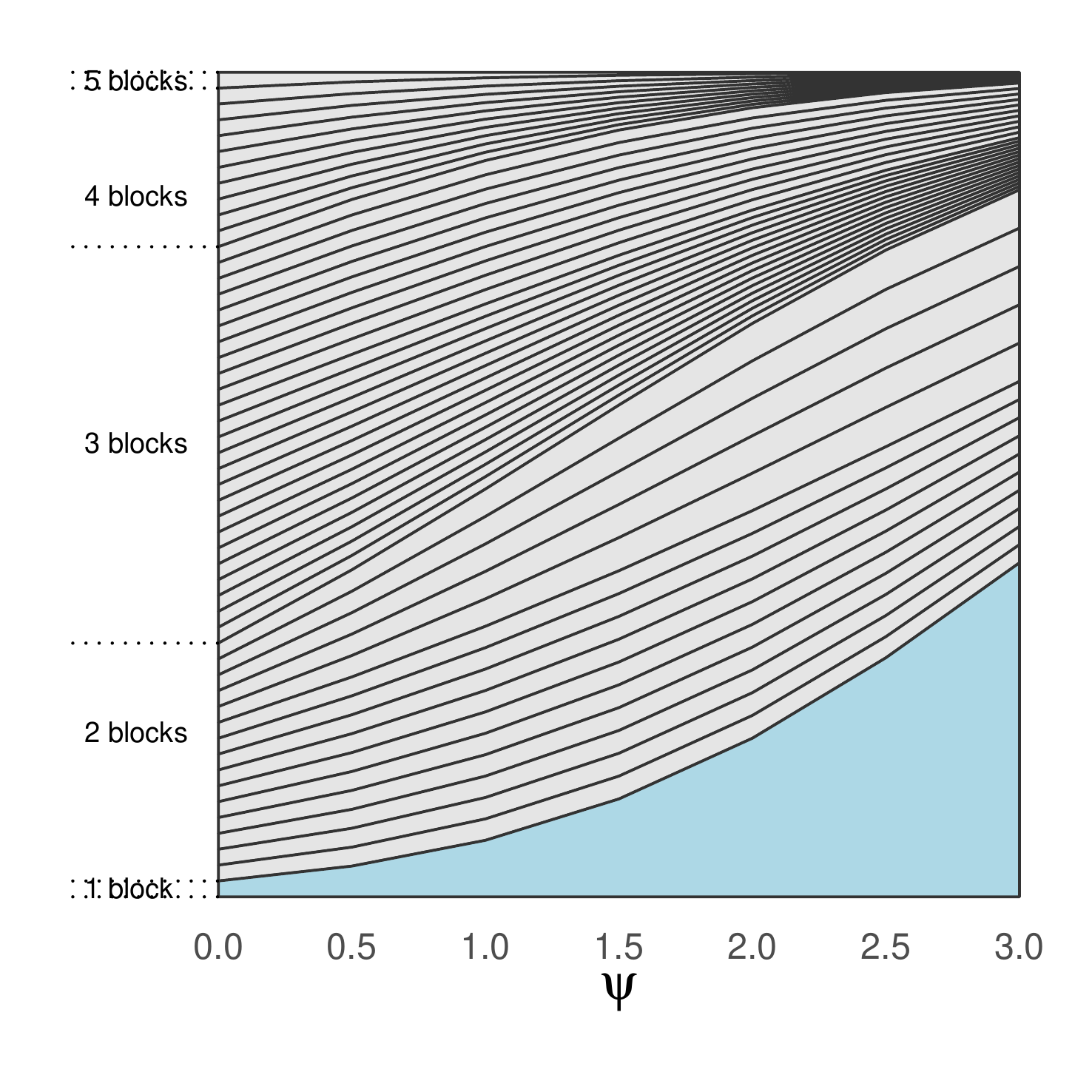}\label{fig:cp_uniform_1}}
 \subfloat[$\boldc_0 = \{1,2\}\{3,4,5\}$]{\includegraphics[width = 0.48\textwidth]{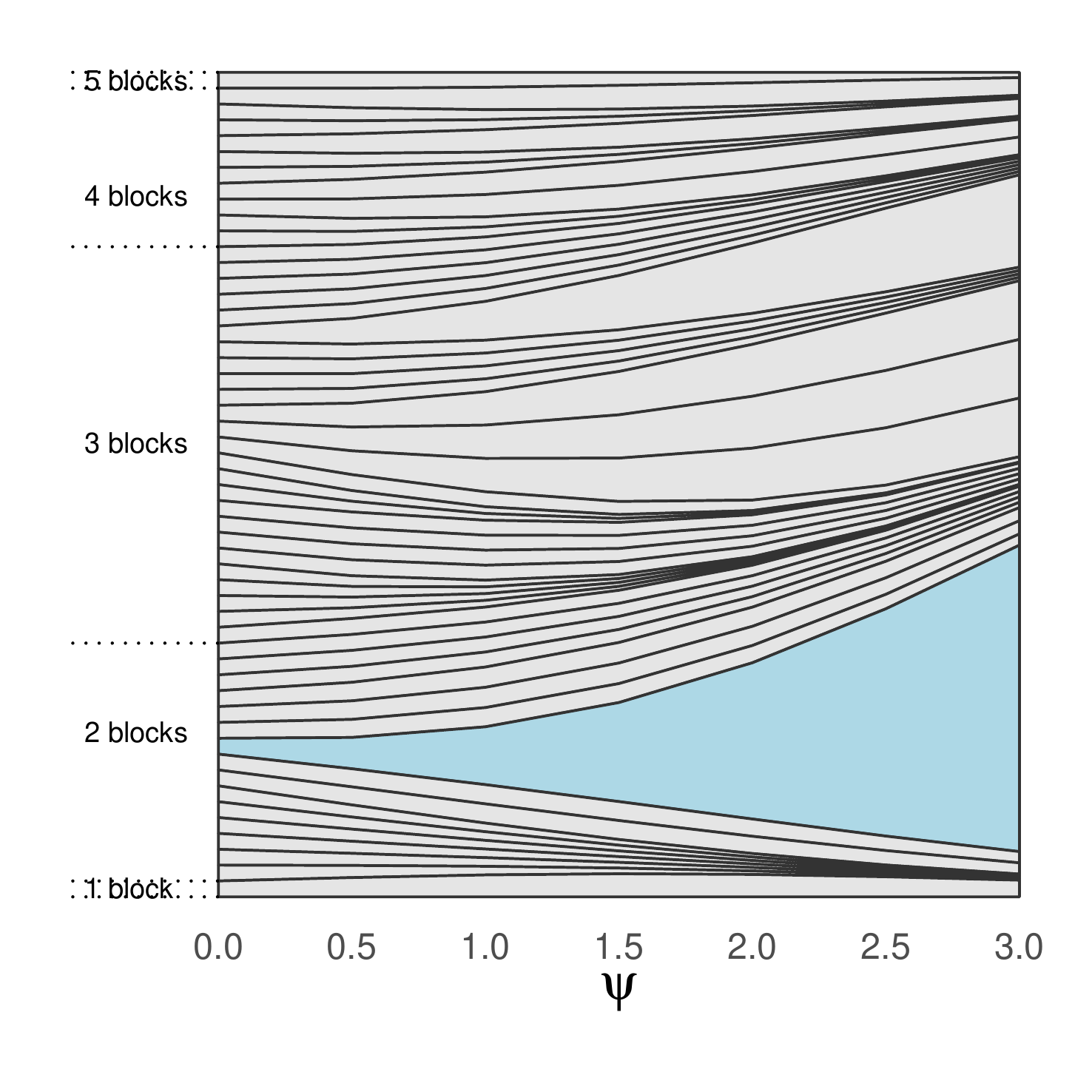}\label{fig:cp_uniform_2}}\\
 \subfloat[$\boldc_0 = \{1,2\}\{3,4\}\{5\}$]{\includegraphics[width = 0.48\textwidth]{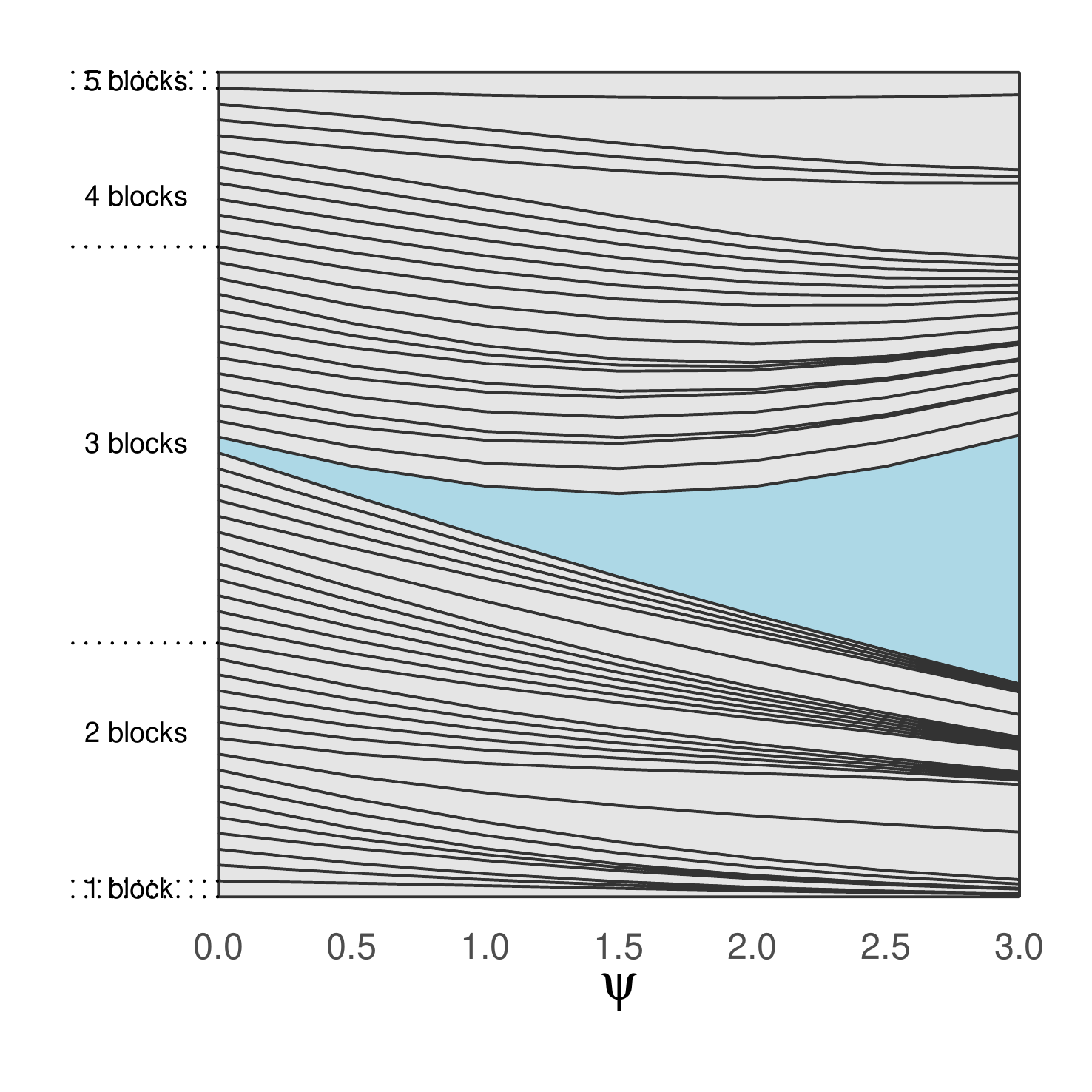}\label{fig:cp_uniform_3}}
 \subfloat[$\boldc_0 = \{1\}\{2\}\{3\}\{4,5\}$]{\includegraphics[width = 0.48\textwidth]{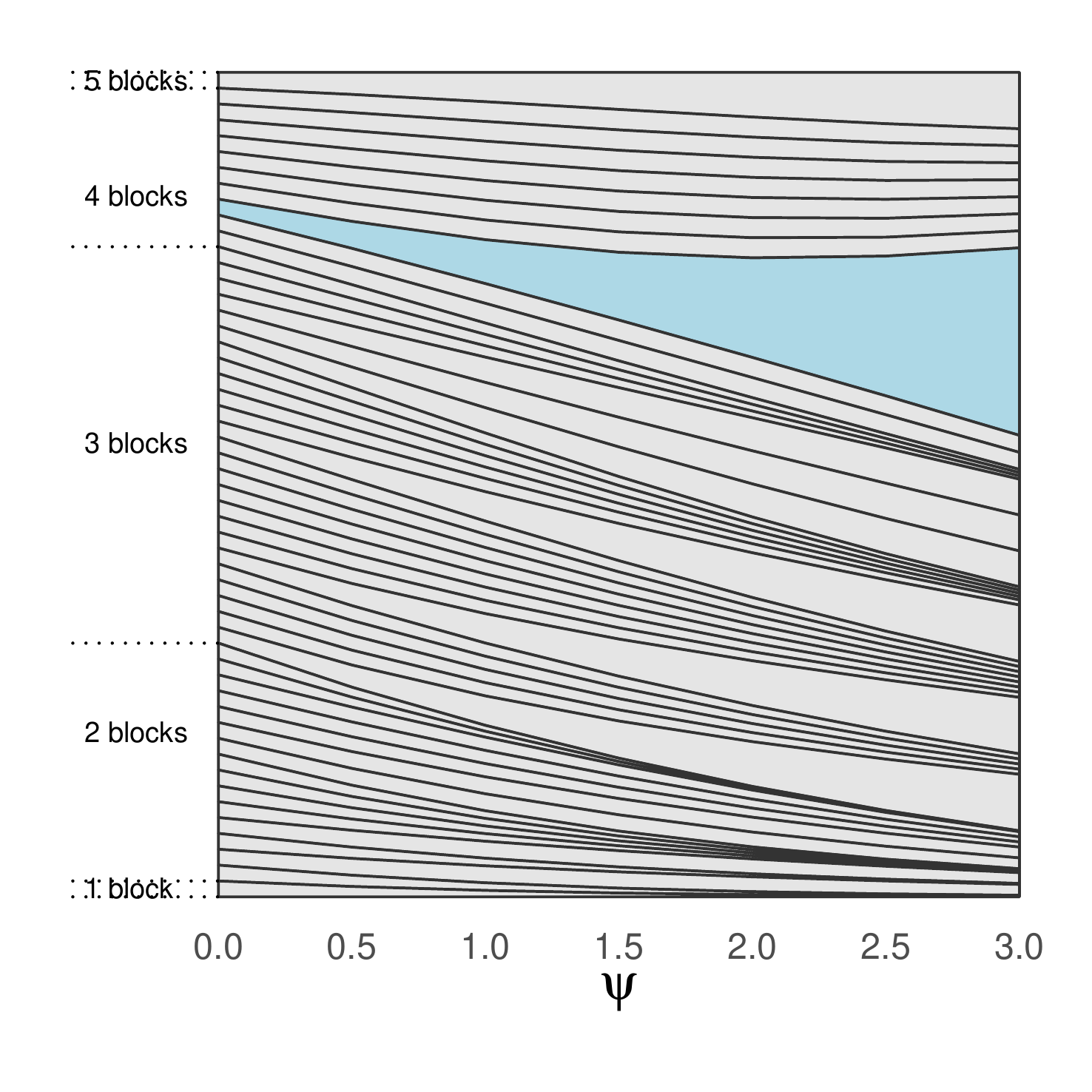}\label{fig:cp_uniform_4}}
 \caption{Prior probabilities of the $52$ set partitions of $N = 5$ elements for the CP process with uniform base EPPF. In each graph the CP process is centered on a different partition $\boldc_0$ highlighted in blue. The cumulative probabilities across different values of the penalization parameter $\psi$ are joined to form the curves, while the probability of a given partition corresponds to the area between the curves.}\label{fig:cp_uniform}
\end{figure}

Considering $N = 5$, there are $52$ possible set partitions; Figure~\ref{fig:cp_uniform} shows the prior probabilities assigned to partitions under the CP process for different values of $\psi \in (0, 3)$ with $\psi = 0$ corresponding to the uniform prior. Notice that base partitions with the same configuration (e.g. for $\boldc_0 = \{1,2\}\{3,4,5\}$ all the partitions with blocks sizes $\{3,2\}$), will behave in the same way, with the same probabilities assigned to partitions different in composition.

\begin{figure}[!ht]
 \subfloat[$\boldc_0 = \{1,2,3,4,5\}$]{\includegraphics[width = 0.48\textwidth]{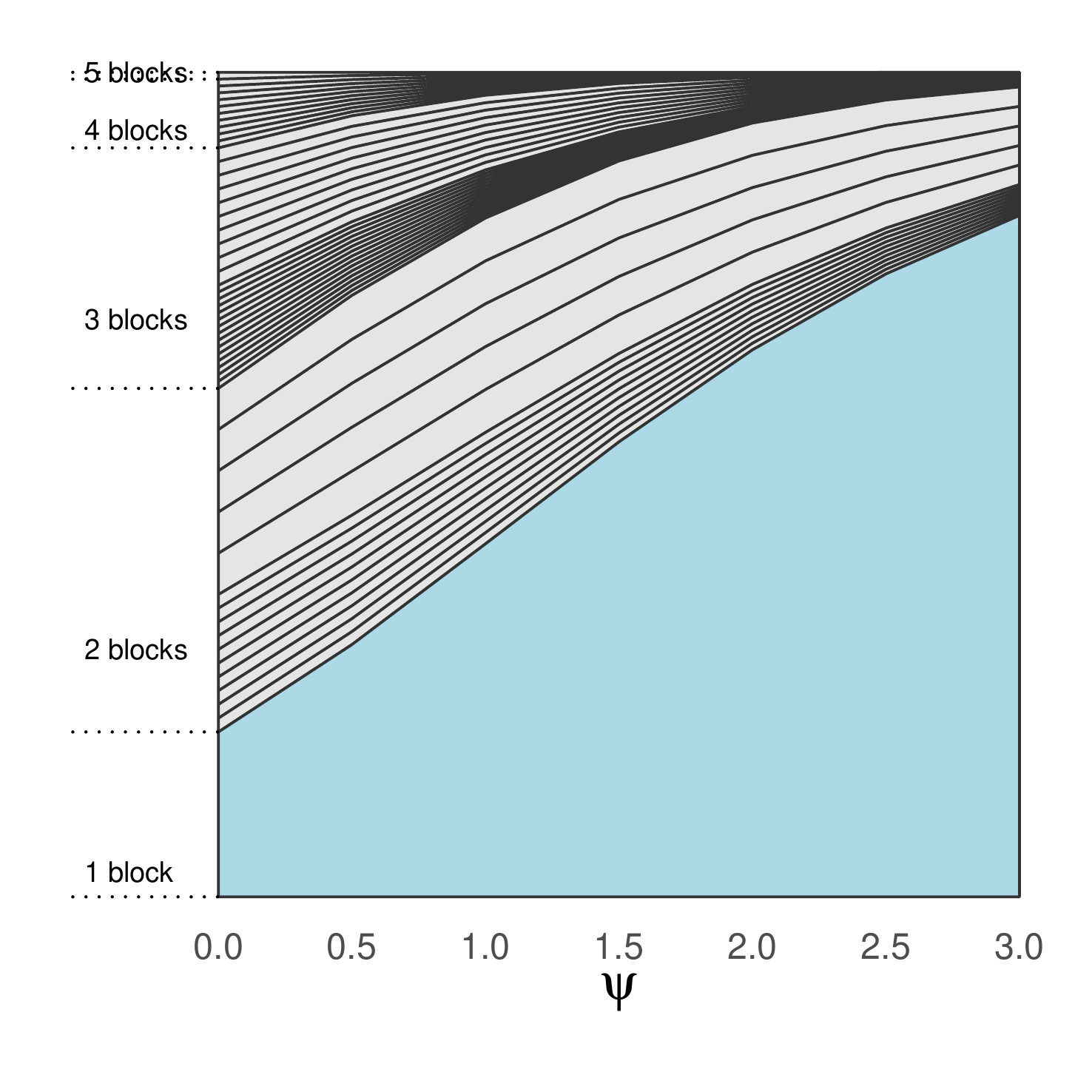}}
 \subfloat[$\boldc_0 = \{1,2\}\{3,4,5\}$]{\includegraphics[width = 0.48\textwidth]{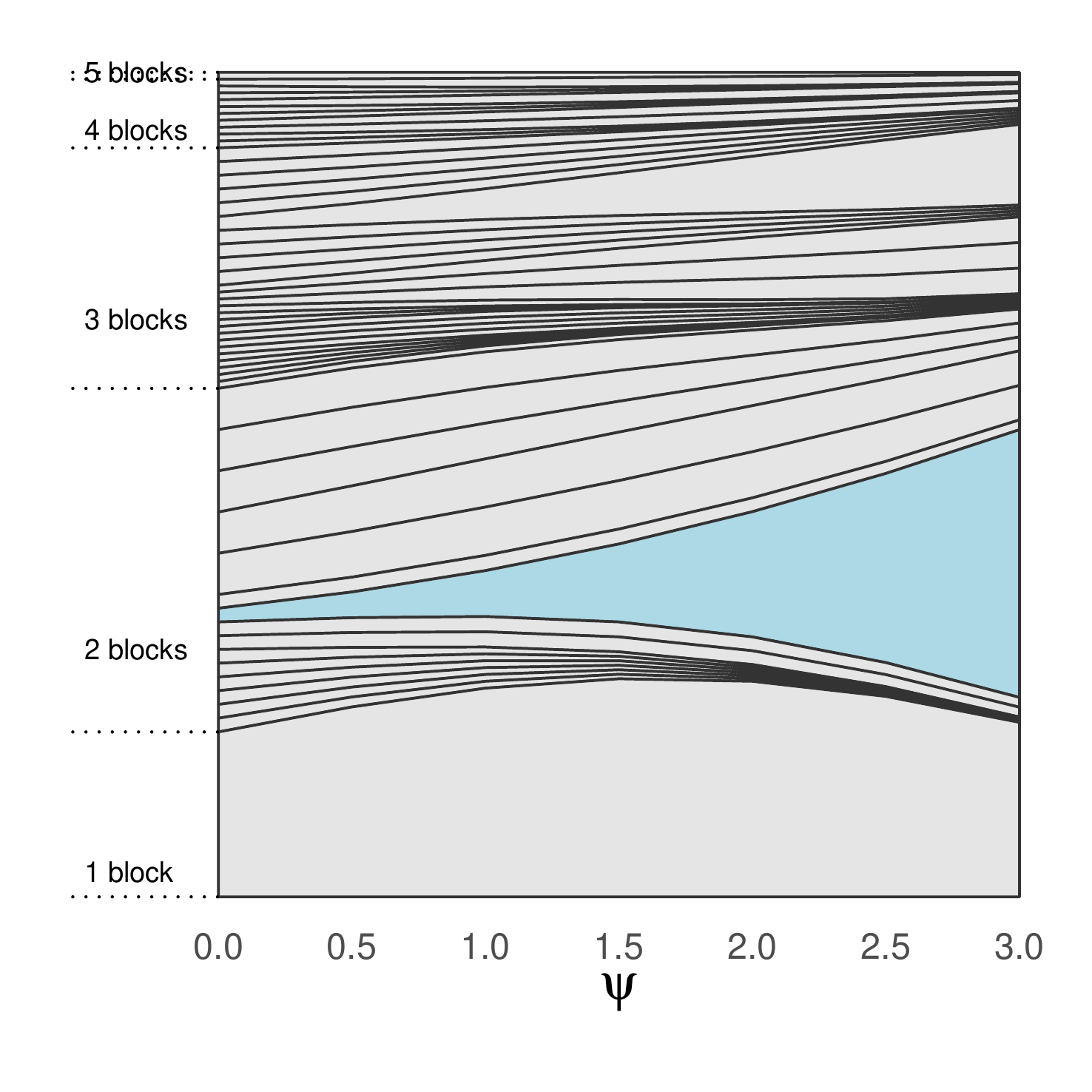}}\\
 \subfloat[$\boldc_0 = \{1,2\}\{3,4\}\{5\}$]{\includegraphics[width = 0.48\textwidth]{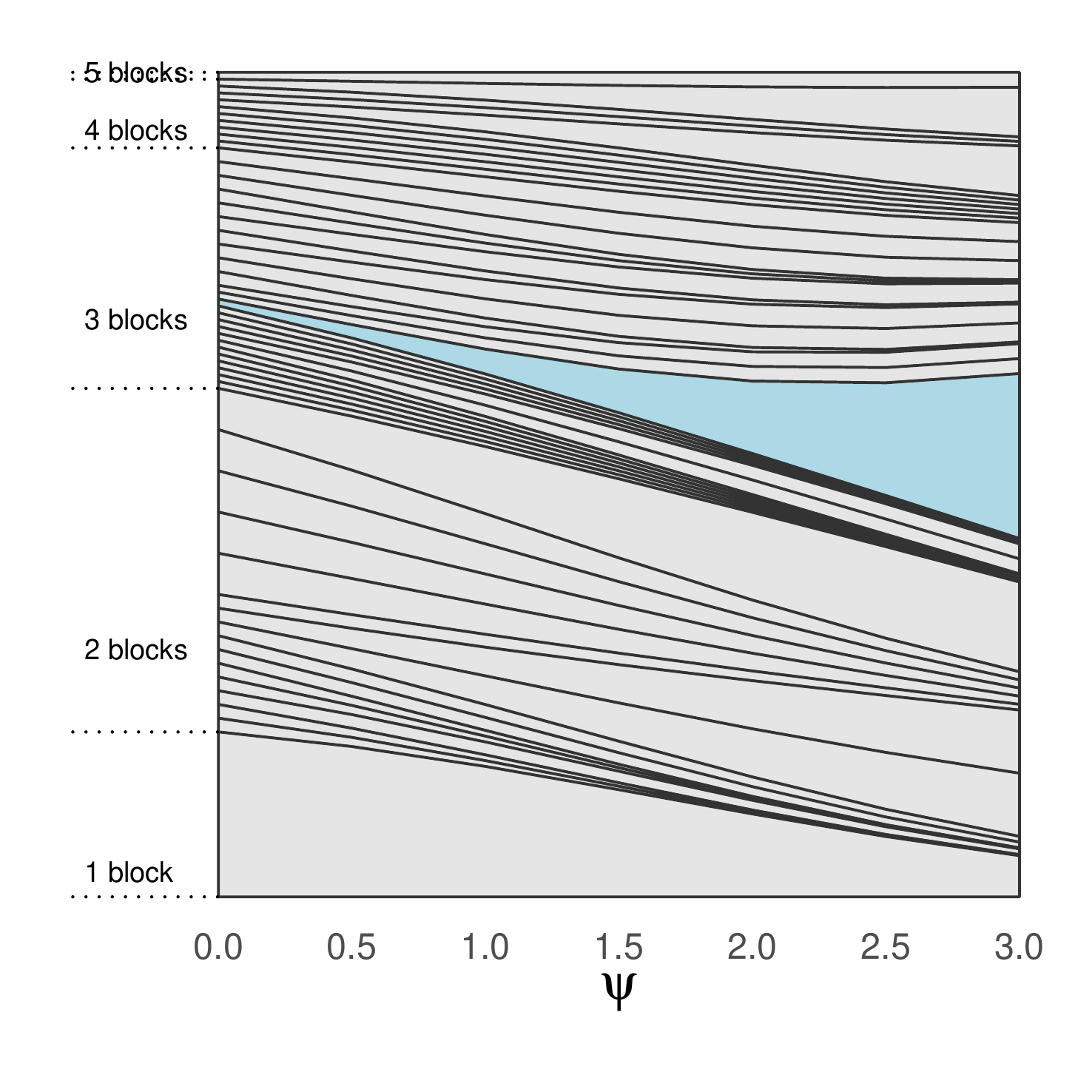}}
 \subfloat[$\boldc_0 = \{1\}\{2\}\{3\}\{4,5\}$]{\includegraphics[width = 0.48\textwidth]{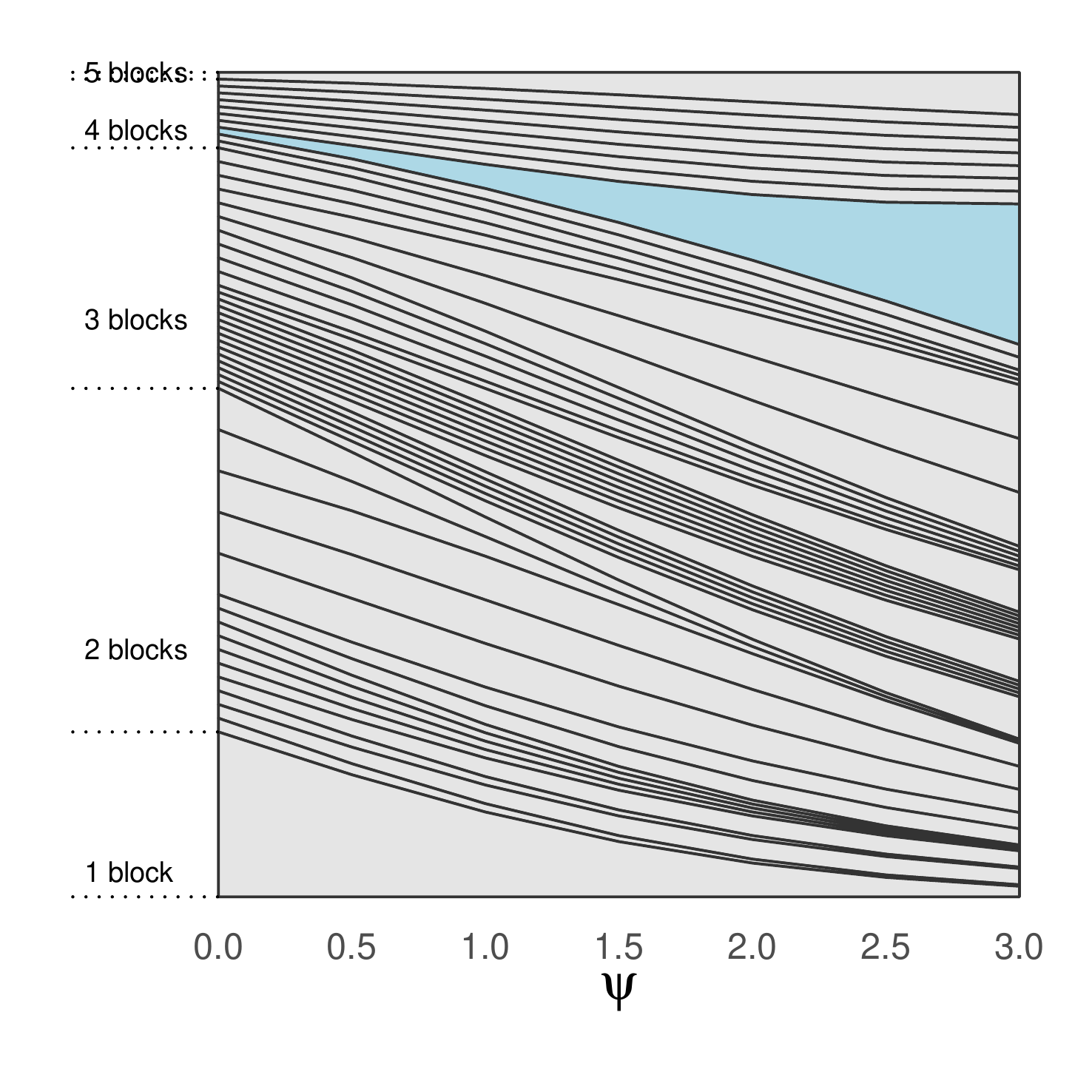}}
 \caption{Prior probabilities of the $52$ set partitions of $N = 5$ elements for the CP process with Dirichlet process of $\alpha = 1$ base EPPF. In each graph the CP process is centered on a different partition $\boldc_0$ highlighted in blue. The cumulative probabilities across different values of the penalization parameter $\psi$ are joined to form the curves, while the probability of a given partition corresponds to the area between the curves.}\label{fig:cp_dp}
\end{figure}

Non-zero values of $\psi$ increase the prior probability of partitions $\boldc$ that are relatively close to the chosen $\boldc_0$. However, the effect is not uniform but depends on the structure of both $\boldc$ and $\boldc_0$.
For example, consider the inflation that occurs in the blue region as $\psi$ increases from $0$ to $3$. When $\boldc_0$ has $2$ blocks (Figure~\ref{fig:cp_uniform_1}) versus $4$
(Figure~\ref{fig:cp_uniform_4}) there is a bigger increase.
This is because the space of set partitions $\Pi_N$ is not ``uniform'', since given a fixed configuration there is a heterogeneous number of partitions.
Rewriting $\boldsymbol{\lambda} = (\lambda_1, \ldots, \lambda_K)$ as $\boldsymbol{\lambda} = (1^{f_1}, 2^{f_2}, \ldots, K^{f_K})$, with the notation indicating that there are $f_i$ elements of $\boldsymbol{\lambda}$ equal to $i$, the number of set partitions with configuration $\boldsymbol{\lambda}$ is
\begin{equation*}\label{eq_integer_enum}
 \frac{N!}{\prod_{j = 1}^K \lambda_j! \prod_{i = 1}^N f_i!}.
\end{equation*}
\noindent For example, for $\{221\} = 1^{1}2^{2}3^{0}4^{0}5^{0}$, the number of corresponding set partitions is $15$, while there are $10$ set partitions of type $\{311\}$.

While the uniform distribution gives the same probability to each partition in the space, the EPPF induced by Gibbs-type priors distinguishes between different configurations, but not among partitions with the same configuration. We focus on the Dirichlet process case, being the most popular process employed in applications. Under the DP the induced EPPF $p_0(\boldc) \propto \alpha^K \prod_{j = 1}^{K} \Gamma(\lambda_j)$ is a function of the configuration $\boldsymbol{\Lambda}(\boldc)$, which is one of $\{\boldsymbol{\lambda}_1, \ldots, \boldsymbol{\lambda}_M\}$ since the possible configurations are finite and correspond to the number of integer partitions. Letting $g(\boldsymbol{\Lambda}(\boldc)) = \alpha^K \prod_{j = 1}^{K} \Gamma(\lambda_j)$, the formulation in~\eqref{eq_cenex} can be written as
\begin{equation}\label{eq:analitic_DP}
 p(\boldc| \boldc_0, \psi) =  \frac{ g(\boldsymbol{\lambda}_m) e^{-\psi \delta_l}}{\sum_{u = 0}^L  \sum_{v = 1}^M n_{uv} g(\boldsymbol{\lambda}_v) e^{-\psi \delta_u} },  \quad \text{for } \boldc \in s_{lm}(\boldc_0),
\end{equation}
where $s_{lm}(\boldc_0) = \{ \boldc \in \Pi_N : d(\boldc, \boldc_0) = \delta_l, \boldsymbol{\Lambda}(\boldc) = \boldsymbol{\lambda}_m\}$, the set of partitions with distance $\delta_l$ from $\boldc_0$ and configuration $\boldsymbol{\lambda}_m$ for $l = 0, 1, \ldots, L$ and $m = 1, \ldots, M$, with $n_{lm}$ indicating the cardinality.
The factorization~\eqref{eq:analitic_DP} applies for the family of Gibbs-type priors in general, with different expressions of $g(\boldsymbol{\Lambda}(\boldc))$.

In Figure~\ref{fig:cp_dp} we consider the prior distribution induced by the CP process when the baseline EPPF $p_0(\boldc)$ comes from a Dirichlet process with concentration parameter $\alpha = 1$, considering the same base partitions and values for $\psi$ as in Figure~\ref{fig:cp_uniform}.
For the same values of the parameter $\psi$, the behavior of the CP process changes significantly due to the effect of the base prior.
In particular, in the top left panel the CP process is centered on $\boldc_0 = \{1, 2, 3, 4, 5\}$, the partition with only one cluster, which is {\em a priori} the most likely one for $\psi = 0$. In general, for small values of $\psi$ the clustering process will most closely resemble that for a DP, and as $\psi$ increases the DP prior probabilities are decreased for partitions relatively far from $\boldc_0$ and increased for $\boldc_0$ relatively close.

\subsection{Posterior computation under Gibbs-type priors}\label{sec4:Posterior computation}

Certain MCMC algorithms for Bayesian nonparametric mixture models can be easily modified for posterior computation in CP process models.
In particular, we adapt the so-called ``marginal algorithms'' developed for Dirichlet and Pitman-Yor processes. These methods are called marginal since the mixing measure $P$ is integrated out of the model and the predictive distribution is used within a MCMC sampler. In the following, we recall Algorithm 2 in \cite{neal2000} and illustrate how it can be adapted to sample from the CP process posterior. We refer to \cite{neal2000} and references therein for an overview and discussion of methods for both conjugate and nonconjugate cases, and to \cite{fall2014} for adaptation to Pitman-Yor processes.

Let $\boldc$ be represented as an $N$-dimensional vector of indices $\{c_1, \ldots, c_N\}$ encoding cluster allocation and let $\theta_k$ be the set of parameters currently associated to cluster $k$.
\sloppy
The prior predictive distribution for a single $c_i$ conditionally on $\boldc^{-i} = \{c_1, \ldots, c_{i-1}, c_{i+1}, \ldots, c_N\}$ is exploited to perform the Gibbs sampling step allocating observations to either a new cluster or one of the existing ones. Algorithm $2$ in \cite{neal2000} updates each $c_i$ sequentially for $i = 1, \ldots, N$ via a reseating procedure, according to the conditional posterior distribution
\begin{equation}\label{eq:polya_update}
 p(c_i = k | \boldc^{-i}, \boldsymbol{\theta}, y_i) \propto
 \begin{cases}
   & p(c_i = k | \boldc^{-i})  p(y_i| \theta_k)  \quad k = 1, \ldots, K^{-}          \\
   & p(c_i = k | \boldc^{-i})  \int p(y_i| \theta) d G_0(\theta) \quad k = K^{-} +1,
 \end{cases}
\end{equation}
with $K^{-}$ the number of clusters after removing observation $i$.
The conditional distribution $p(c_i = k | \boldc^{-i})$ is reported in Table~\ref{tab:eppf_conditional} for different choices of the prior EPPF. Notice that, for the case of finite Dirichlet prior, the update consists only in the first line of equation~\eqref{eq:polya_update}, since the number of classes is fixed.
For Dirichlet and Pitman-Yor processes, when observation $i$ is associated to a new cluster, a new value for $\theta$ is sampled from its posterior distribution based on the base measure $G_0$ and the observation $y_i$. This approach is straightforward when we can compute the integral $\int p(y_i| \theta) d G_0(\theta)$, as will generally be the case when $G_0$ is a conjugate prior.

\begin{table}
 \centering
 $\begin{array}{ lcc }
   \toprule
   \text{Random probability measure} & \text{Parameters} & p(c_i = k| \boldc^{-i}) \propto \\
   \midrule
   \text{Dirichlet process}          & (\alpha)          &
   \begin{cases}
     & \frac{\lambda_k^{-i}}{\alpha + N - 1} \quad k = 1, \ldots, K^{-} \\
     & \frac{\alpha}{\alpha + N - 1} \quad k = K^{-} +1
   \end{cases}                                                              \\
   \text{Pitman-Yor process}                 & (\alpha, \sigma)  &
   \begin{cases}
     & \frac{\lambda_k^{-i} - \sigma}{\alpha + N - 1} \quad k = 1, \ldots, K^{-} \\
     & \frac{\alpha + \sigma K^{-}}{\alpha + N - 1} \quad k = K^{-} +1
   \end{cases}
   \\
   \midrule
   \text{Symmetric Dirichlet}        & (\kappa, \gamma)  &
   \frac{\lambda_k^{-i} + \gamma/\kappa}{\alpha + N - 1} \quad k = 1, \ldots, \kappa       \\
   \bottomrule
  \end{array}$
 \caption[Conditional prior distribution for $c_i$ given $\boldc^{-i}$ under different choices of the EPPF.]{Conditional prior distribution for $c_i$ given $\boldc^{-i}$ under different choices of the EPPF. With $K^{-}$ we denote the total number of clusters after removing the $i$th observation while $\lambda_k^{-i}$ is the corresponding size of cluster $k$.}
 \label{tab:eppf_conditional}
\end{table}

Considering the proposed CP process, the conditional distribution for $c_i$ given $\boldc^{-i}$ can still be computed, but it depends both on the base prior and the penalization term accounting for the distance between the base partition $\boldsymbol{c}_0$ and the one obtained by assigning the observation $i$ to either one of the existing classes $k \in\{1, \ldots, K^{-}\}$ or a new one.
Hence, the step in equation~\eqref{eq:polya_update} can be easily adapted by substituting the conditional distribution for $p(c_i = k | \boldc^{-i})$ with
\begin{equation*}\label{eq:conditional_cp}
 p(c_i = k | \boldc^{-i}, \boldc_0, \psi ) \propto p_0(c_i = k | \boldc^{-i}) \exp\{ - \psi d(\boldc, \boldc_0) \} \quad k = 1, \ldots, K^{-}, K^{-} +1
\end{equation*}
with $\boldc = \{\boldc^{-i} \cup \{c_i= k\}\}$ the current state of the clustering and $p_0(c_i = k | \boldc^{-i})$ one of the conditional distributions in Table~\ref{tab:eppf_conditional}. Additional steps  on the implementation using the variation of information as a distance are given in the Appendix (Algorithm~\ref{alg:distance}).

Extension to the non-conjugate context can be similarly handled exploiting Algorithm $8$ in \cite{neal2000} based on auxiliary parameters, which avoids the computation of the integral $\int p(y_i| \theta) d G_0(\theta)$. The only difference is that, when $c_i$ is updated, $m$ temporary auxiliary variables are introduced to represent possible values of components parameters that are not associated with any other observations. Such variables are simply sampled from the base measure $G_0$, with the probabilities of a new cluster in Table~\ref{tab:eppf_conditional} changing into $(\alpha/m)/(\alpha + N - 1)$ for the Dirichlet process and to $[(\alpha + \sigma K^{-})/m]/(\alpha + N - 1)$ for the Pitman-Yor process, for $k = K^{-} + 1, \ldots,K^{-} + m$.

\section{Prior calibration}\label{sec:prior_calibration}
As the number of observations $N$ increases, the number of partitions explodes, and higher values of $\psi$ are needed to place non-negligible prior probability in small to moderate neighborhoods around $\boldc_0$.
The prior concentration around $\boldc_0$ depends on three main factors: i) $N$ through $\mathcal{B}_N$, i.e. the cardinality of the space of set partitions, ii) the baseline EPPF $p_0(\boldc_0)$ and iii) where $\boldc_0$ is located in the space. We hence propose a general method to evaluate the prior behavior under different settings, while suggesting how to choose the parameter $\psi$.

One may evaluate the prior distribution for different values of $\psi$ and check its behavior using graphs such as those in Section~\ref{sec:3.3_prior}, however they become difficult to interpret as the space of partitions grows.
We propose to evaluate the probability distribution of the distances $\delta = d(\boldc, \boldc_0 )$ from the known partition $\boldc_0$. The probability assigned to different distances by the prior is
\begin{equation*}
 p(\delta = \delta_l) = \sum_{\boldc \in \Pi_N} p(\boldc)\mathcal{I} \left\{ d(\boldc,\boldc_0) =\delta_l)\right\} = \sum_{\boldc \in s_l(\boldc_0)} p(\boldc) \quad l = 0, \dots, L,
\end{equation*}
with $\mathcal{I}(\cdot)$ the indicator function and $s_l(\boldc_0)$ denoting the set of partitions distance $\delta_l$ from $\boldc_0$, as defined in \eqref{eq:set_c0}.
Consider the uniform distribution on set partitions, $p(\delta = \delta_l) = |s_l(\boldc_0)|/\mathcal{B}_N$, the proportion of partitions distance $\delta_l$ from $\boldc_0$. Under the general definition of the CP process, the resulting distribution becomes
\begin{equation}\label{eq:distance_probabilities}
 p(\delta = \delta_l) = \sum_{\boldc \in s_l(\boldc_0)} \frac{p_0(\boldc) e^{-\psi \delta_l}}{\sum_{u = 0}^L \sum_{\boldc^* \in s_u(\boldc_0)} p_0(\boldc^*) e^{-\psi \delta_u }} \quad l = 0, \ldots, L,
\end{equation}
with the case of Gibbs-type EPPF corresponding to
\begin{equation}\label{eq:distance_probabilities_dp}
 p(\delta = \delta_l) =  \frac{ \sum_{m = 1}^M n_{lm} g(\boldsymbol{\lambda}_m) e^{-\psi \delta_l}}{\sum_{u = 0}^L  \sum_{v = 1}^M n_{uv} g(\boldsymbol{\lambda}_v) e^{-\psi \delta_u} }, \quad l = 0, \ldots, L.
\end{equation}
\noindent
Notice that the uniform EPPF case is recovered when $g(\boldsymbol{\lambda}_m) = 1$ for $m = 0, \ldots, M$, so that $\sum_{m = 1}^M n_{lm} = n_l$. Hence the probability in~\eqref{eq:distance_probabilities} simplifies to
\begin{equation}\label{eq:distance_probabilities_uniform}
 p(\delta = \delta_l) =  \frac{n_l e^{-\psi \delta_l}}{\sum_{u = 0}^L n_u e^{-\psi \delta_u}} \quad l = 0, \ldots, L.
\end{equation}
\noindent
In general, since distances are naturally ordered, the corresponding cumulative distribution function can be simply defined as $F(\delta) = \sum_{\delta_l \leq \delta} p(\delta_l)$ for $\delta \in \{\delta_0, \ldots, \delta_L\}$ and used to assess how much mass is placed in different size neighborhoods around $\boldc_0$ under different values of $\psi$.
Hence we can choose $\psi$ to place a specified probability $q$ (e.g. $q = 0.9$) on partitions within a specified distance $\delta^*$ from $\boldc_0$. This would correspond to calibrating $\psi$ so that $F(\delta^*) \approx q$, with $F(\delta^*) \geq q$. In other words, partitions generated from the prior would have at least probability $q$ of being within distance $\delta^*$ from $\boldc_0$.

The main problem is in computing the probabilities in equations~\eqref{eq:distance_probabilities_dp}-\eqref{eq:distance_probabilities_uniform}, which depend on all the set partitions in the space. In fact, one needs to count all the partitions having distance $\delta_l$ for $l = 0, \ldots, L$ when the base EPPF is uniform, while taking account of configurations in the case of the Gibbs-type priors.
Even if there are quite efficient algorithms to list all the possible set partitions of $N$ \citep[see][]{knuth2005art, nijenhuis2014}, it becomes computationally infeasible due to the extremely rapid growth of the space; for example from $N = 12$ to $13$, the number of set partitions grows from $\mathcal{B}_{12} = 4,213,597$ to $\mathcal{B}_{13}= 27,644,437$.

We propose a general strategy to approximate prior probabilities assigned to different distances from $\boldc_0$ focused on obtaining estimates of distance values and related counts, which represent the sufficient quantities to compute \eqref{eq:distance_probabilities_dp}-\eqref{eq:distance_probabilities_uniform} under different values of $\psi$. We consider a targeted Monte Carlo procedure which augments uniform sampling on the space of set partitions with a deterministic local search using the Hasse diagram to estimate the counts for small values of the distance.
\subsection{Deterministic local search}
\noindent
Poset theory provides a nice representation of the space of set partitions by means of the Hasse diagram illustrated in Section~\ref{sec2.1}, along with suitable definition of metrics. A known partition $\boldc_0$ can be characterized in terms of number of blocks $K_0$ and configuration $\boldsymbol{\Lambda}(\boldc_0)$. These elements allows one to locate $\boldc_0$ in the Hasse diagram and then explore connected partitions by means of split and merge operations on the clusters in $\boldc_0$.

\begin{figure}[!tb]
 \centering
 \begin{tikzpicture}[align = center, scale = 0.72, transform shape]
  \node (1) at (0,0) {$\textcolor{MidnightBlue}{\textbf{\{1,2,3,4\}}}$};
  \node [below =1cm of 1] (2) {$\textcolor{JungleGreen}{\textbf{\{4\}\{1,2,3\}}}$};
  \node [left=0.6cm of 2] (3)  {$\textcolor{JungleGreen}{\textbf{\{3\}\{1,2,4\}}}$};
  \node [left=0.6cm of 3] (4)  {$\textcolor{JungleGreen}{\textbf{\{2\}\{1,3,4\}}}$};
  \node [left=0.6cm of 4] (5)  {$\textbf{\{1\}\{2,3,4\}}$};

  \node [right =0.6cm of 2] (6) {$\textcolor{JungleGreen}{\textbf{\{1,2\}\{3,4\}}}$};
  \node [right =0.6cm of 6] (7) {$\textcolor{JungleGreen}{\textbf{\{1,3\}\{2,4\}}}$};
  \node [right =0.6cm of 7] (8) {$\textcolor{JungleGreen}{\textbf{\{1,4\}\{2,3\}}}$};

  \node [below =1cm of 5] (9) {$\textcolor{MidnightBlue}{\textbf{\{1\}\{2\}\{3,4\}}}$};
  \node [right =0.9cm of 9] (10) {$\textcolor{MidnightBlue}{\textbf{\{1\}\{3\}\{2,4\}}}$};
  \node [right =0.9cm of 10] (11) {$\textcolor{MidnightBlue}{\textbf{\{1\}\{4\}\{2,3\}}}$};
  \node [right =0.9cm of 11] (12) {$\textcolor{SkyBlue}{\textbf{\{2\}\{3\}\{1,4\}}}$};
  \node [right =0.9cm of 12] (13) {$\textcolor{SkyBlue}{\textbf{\{2\}\{4\}\{1,3\}}}$};
  \node [right =0.9cm of 13] (14) {$\textcolor{SkyBlue}{\textbf{\{3\}\{4\}\{1,2\}}}$};
  \node [below =3cm of 2] (0) {$\textcolor{JungleGreen}{\textbf{\{1\}\{2\}\{3\}\{4\}}}$};

  \draw [gray, densely dashed, line width=0.25mm] (1) -- (2) (1) -- (3) (1) -- (4) (1) -- (6) (1) -- (7) (1) -- (8);
  \draw [white!30!black,  thick, line width=0.25mm] (1) -- (5);
  \draw [white!30!black,  thick, line width=0.25mm] (5) -- (9) (5) -- (10) (5) -- (11);
  \draw [white!40!black,  densely dashed, line width=0.25mm] (6) -- (9) (7) -- (10) (8) -- (11) (9) -- (0) (10) -- (0) (11) -- (0) (4) -- (9) (3) -- (10) (2) -- (11);

  \draw [gray,  dotted, line width=0.25mm]  (4) -- (12) (4) -- (13);
  \draw [gray,  dotted, line width=0.25mm]  (3) -- (12) (3) -- (14);
  \draw [gray,  dotted, line width=0.25mm]  (2) -- (13) (2) -- (14);
  \draw [white!40!black,  dotted, line width=0.25mm] (6) -- (14) (7) -- (13) (8) -- (12);

  \draw [gray,  dotted, line width=0.25mm]  (12) -- (0) (13) -- (0) (14) -- (0);

 \end{tikzpicture}
 \caption[Local search example on $\Pi_4$.]{Illustration of results from the local search algorithm based on the Hasse diagram of $\Pi_4$ starting from $\boldc_{0} = \textbf{\{1\}\{2,3,4\}}$. Partitions are colored according the exploration order according to dark-light gradient. Notice that after $3$ iterations the space is entirely explored.}
 \label{fig:hasse_search}
\end{figure}
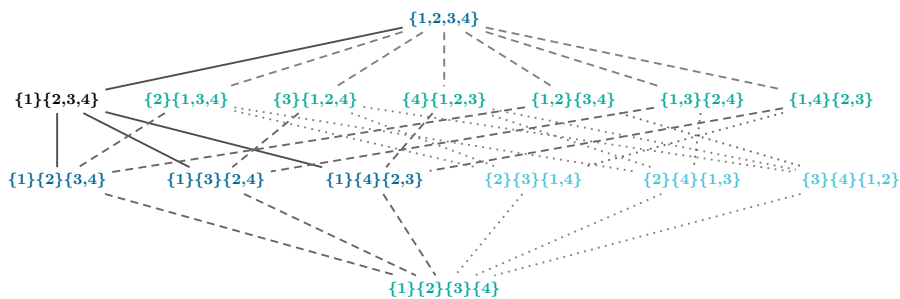

As an illustrative example, consider the Hasse diagram of $\Pi_4$ in Figure~\ref{fig:hasse_search} and $\boldc_0 =  \{1\}\{2,3,4\}$, having $2$ clusters and configuration $\boldsymbol{\Lambda}({\boldc_0}) = \{31\}$.
Let $\mathcal{N}_1(\boldc_0)$ denote the sets of partitions directly connected with $\boldc_0$, i.e. partitions covering $\boldc_0$ and those covered by $\boldc_0$. In general, a partition $\boldc_0$ with $K_0$ clusters is covered by ${K_0}\choose{2}$ partitions and covers $\sum_{j = 1}^{K_0} 2^{\lambda_j - 1} - 1$. In the example, $\mathcal{N}_1(\boldc_0)$ contains $\{1,2,3,4\}$ obtained from $\boldc_0$ with a merge operation on the two clusters, and all the partitions obtained by splitting the cluster $\{2,3,4\}$ in any possible way.
The base idea underlying the proposed local search, consists in exploiting the Hasse diagram representation to find all the partitions in increasing distance neighborhoods of $\boldc_0$. One can list partitions at $T$ connections from $\boldc_0$ starting from $\mathcal{N}_1(\boldc_0)$ by recursively applying split and merge operations on the set of partitions explored at each step. Potentially, with enough operations one can reach all the set partitions, since the space is finite with lower and upper bounds.

In practice, the space is too huge to be explored entirely, and a truncation is needed. From the example in Figure~\ref{fig:hasse_search}, $\mathcal{N}_1(\boldc_0)$ contains $3$ partitions with distance $0.69$ from $\boldc_0$ and one with distance $1.19$. Although $\mathcal{N}_2(\boldc_0)$ may contain partitions closer to $\boldc_0$ than this last, the definition of distance in Section~\ref{sec3:distance} guarantees that there are no other partitions with distance from $\boldc_0$ less than $0.69$.
Since the VI is the minimum weighted path between two partitions, all the partitions reached at the second exploration step add a nonzero weight to distance computation. This consideration extends to an arbitrary number of explorations $T$, with $\delta_{L^*} = \text{min}\{d(\boldc^*, \boldc_0)\}_{\boldc^* \in \mathcal{N}_T(\boldc_0)}$ being the upper bound on the distance value. By discarding all partitions with distance greater that $\delta_{L^*}$, one can compute exactly the counts in equations~\eqref{eq:distance_probabilities_dp}-\eqref{eq:distance_probabilities_uniform} related to distances $\delta_0, \ldots, \delta_{L^*}$. Notice that $2/N$ is the minimum distance between two different partitions, and $2T/N$ is a general lower bound on the distances from $\boldc_0$ that can be reached in $T$ iterations.

\subsection{Monte Carlo approximation}\label{sec:prior_monte_carlo}
\noindent
We pair the local exploration with a Monte Carlo procedure to estimate the counts and distances greater that $\delta_{L^*}$, in order to obtain a more refined representation of the prior distance probabilities. Sampling uniformly from the space of partitions is not in general a trivial problem, but a nice strategy has been proposed in \cite{stam1983}, in which the probability of a partition with $K$ clusters is used to sample partitions via an urn model. Derivation of the algorithm starts from the \textit{Dobi{\'n}ski formula} \citep{dobinski1877} for the Bell numbers
\begin{equation}\label{eq:dobinsky}
 \mathcal{B}_N = e^{-1} \sum_{k = 1}^{\infty} \frac{k^N}{k!},
\end{equation}
which from a probabilistic perspective corresponds to the $k$-th moment of the Poisson distribution with expected value equal to $1$. Then a probability distribution for the number of clusters $K \in \{1, \ldots, N\}$ of a set partition can be defined as
\begin{equation}\label{eq:prob_number_clusters}
 P(K = k) = e^{-1} \frac{k^N}{\mathcal{B}_N k!},
\end{equation}
which is a well defined law thanks to \eqref{eq:dobinsky}. To simulate a uniform law over $\Pi_N$, \cite{stam1983}'s algorithm first generates the number of clusters $K$ according to \eqref{eq:prob_number_clusters} and, conditionally on the sampled value, it allocates observations to the clusters according a discrete uniform distribution over $\{1, \ldots, K\}$. We refer to \cite{stam1983} and \cite{pitman1997setpart} for derivations and proof of the validity of the algorithm.

We adapt the uniform sampling to account for the values already computed by rejecting all the partitions with distance less that $\delta_{L^*}$, restricting the space to $\left\{\Pi_N \setminus \{\mathcal{N}_t(\boldc_0)\}_{t = 0}^T \right\}$.
In practice, few samples are discarded since the probability to sample one such partition corresponds to $|\{\mathcal{N}_t(\boldc_0)\}_{t = 0}^T|/\mathcal{B}_N$, which is negligible for small values of exploration steps $T$ that are generally used in the local search.
A sample of partitions $\boldc^{(1)}, \ldots, \boldc^{(R)}$, can be used to provide an estimate of the counts. Let $R^*$ denote the number of accepted partitions and $\mathcal{B}^*= \mathcal{B}_N - |\{\mathcal{N}_t(\boldc_0)\}_{t = 0}^T|$ be the number of partitions in the restricted space. Conditionally on the observed values of distances in the sample, $\hat{\delta}_{(L^* + 1)}, \ldots, \hat{\delta}_L$, an estimate of the number of partitions with distance $\hat{\delta}_l$ to use in the uniform EPPF case is
\begin{equation}\label{eq:est_nl}
 \hat{n}_l = \mathcal{B}^* \frac{1}{R^*} \sum_{r = 1}^{R^*} \mathcal{I}\left\{ d(\boldc^{(r)}, \boldc_0) = \hat{\delta_l} \right\},
\end{equation}
obtained by multiplying the proportions of partitions in the sample by the total known number of partitions. For the Gibbs-type EPPF case one needs also to account for the configurations $\boldsymbol{\lambda}_1, \ldots, \boldsymbol{\lambda}_M$ in a given orbital of the distance; hence, the estimates are
\begin{equation}\label{eq:est_nlm}
 \hat{n}_{lm} = \mathcal{B}^* \frac{1}{R^*} \sum_{r = 1}^{R^*} \mathcal{I}\left\{ d(\boldc^{(r)}, \boldc_0) = \hat{\delta_l} \right\}  \mathcal{I}\left\{ \boldsymbol{\Lambda}(\boldc^{(r)}) = \boldsymbol{\lambda}_m \right\}.
\end{equation}
\noindent
Pairing these estimates with the counts obtained via the local search, one can evaluate the distributions in equations~\eqref{eq:distance_probabilities_dp}-\eqref{eq:distance_probabilities_uniform} for different values of $\psi$.
The entire procedure is summarized in Algorithm~\ref{alg:prior_est} in the Appendix. Although it requires a considerable number of steps, the procedure can be performed one single time providing information for different choices of $\psi$ and EPPFs. Moreover the local search can be implemented in parallel to reduce computational costs.

\begin{figure}[!htb]
 \subfloat{\includegraphics[width= 0.5\textwidth]{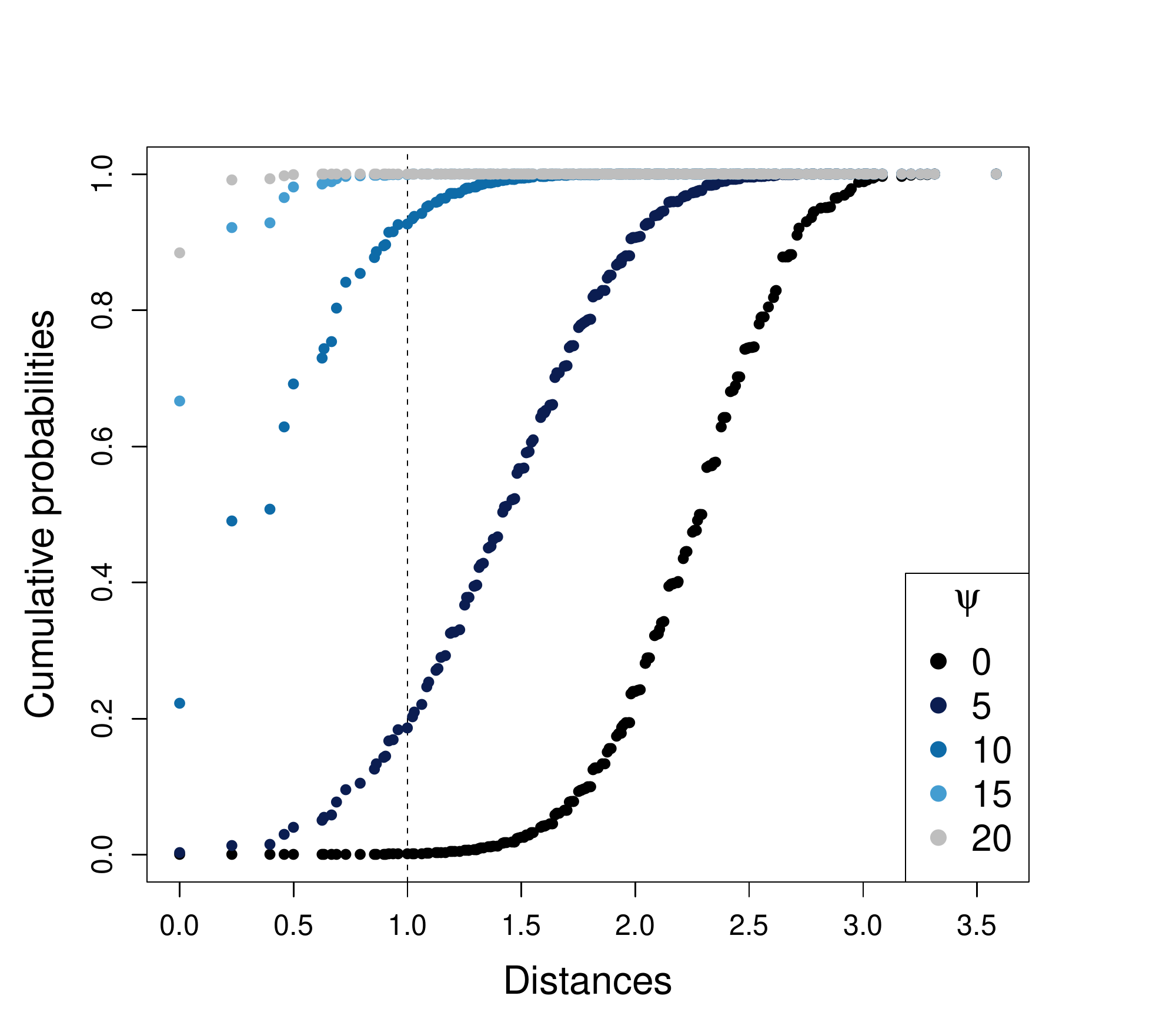}}\hfill
 \subfloat{\includegraphics[width= 0.5\textwidth]{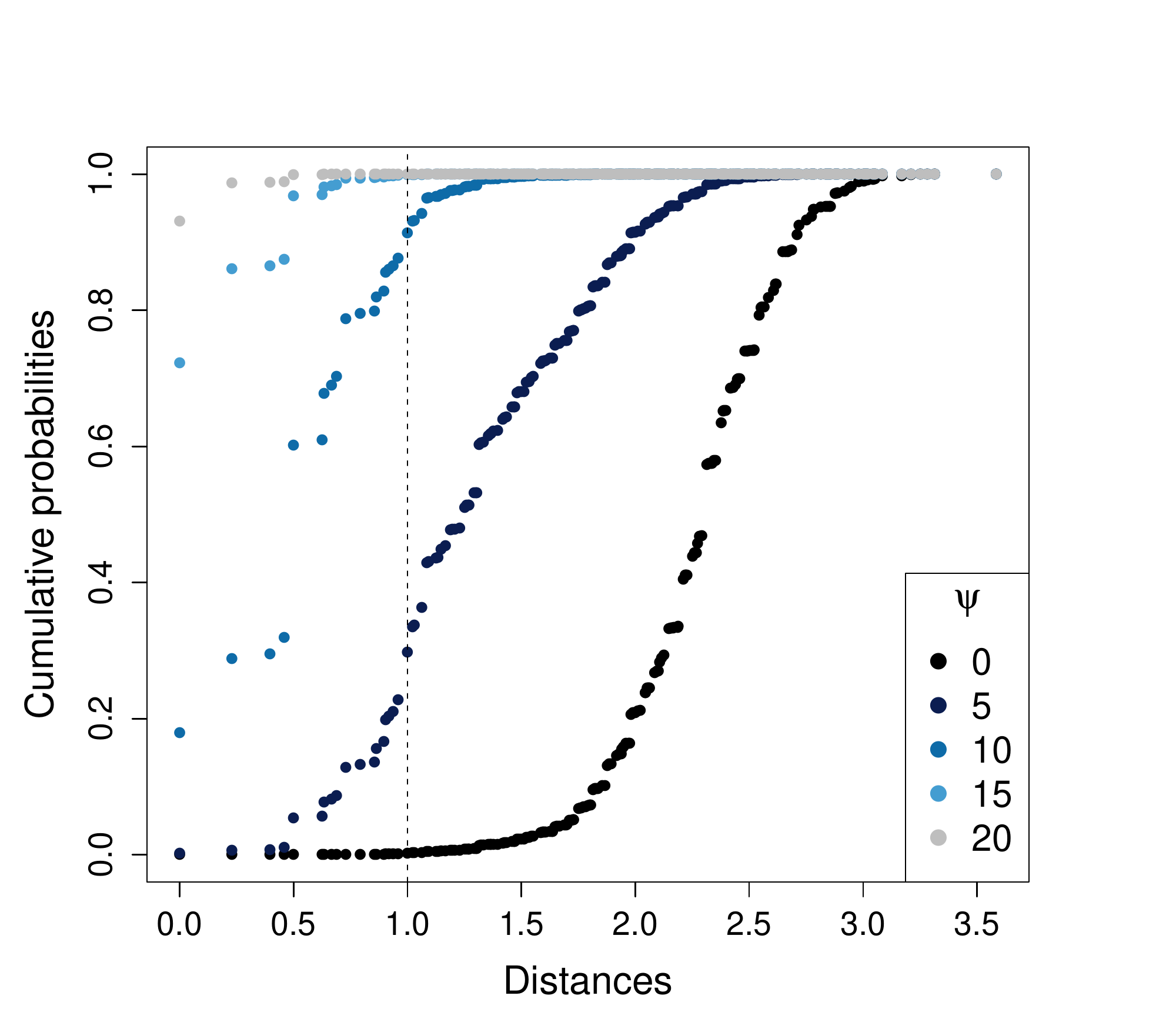}} \\
 \subfloat[Uniform EPPF]{
\resizebox{0.45\textwidth}{!}{%
 \begin{tabular}{r|ccccc}
   \hline
   $\psi$                       & 0    & 5    & 10   & 15   & 20 \\
   min $\hat{\delta}: F(\hat{\delta}) \geq 0.9$ & 2.68 & 1.97 & 0.90 & 0.5 & 0.22  \\
   \hline
  \end{tabular}%
}
 }\hfill
 \subfloat[DP ($\alpha = 1$) EPPF]{
 \resizebox{0.45\textwidth}{!}{%
  \begin{tabular}{r|ccccc}
  \hline
   $\psi$                       & 0    & 5    & 10   & 15  & 20 \\
   min $ \hat{\delta}: F(\hat{\delta}) \geq 0.9$ & 2.71 & 2.04 & 1.27 & 0.5 & 0  \\
  \hline
  \end{tabular}%
  }
 }
 \caption{Estimates of the cumulative prior probabilities assigned to different distances from $\boldc_0$ for $N = 12$ and $\boldc_0$ with configuration $\{3,3,3,3\}$, under the CP process with uniform prior on the left and Dirichlet process on the right. Black dots correspond to the base prior with no penalization, while dots from bottom-to-top correspond to increasing values of $\psi \in \{5, 10, 15, 20\}$. Tables report the minimum distance values such that $F(\delta) \geq 0.9$.}
 \label{fig:est_n12}
\end{figure}

We consider an example for $N = 12$ and $\boldc_0$ with configuration $\{3,3,3,3\}$. Figure~\ref{fig:est_n12} shows the resulting cumulative probability estimates of the CP process under uniform and DP($\alpha = 1$) base distributions, estimated with $m = 4$ iterations of the local search and $20,000$ samples. Dots represent values of the cumulative probabilities, with different colors in correspondence to different values of the parameter $\psi$. Using these estimates one can assess how much probability is placed in different distance neighborhoods of $\boldc_0$; tables in Figure~\ref{fig:est_n12} show the distance values defining neighborhoods around $\boldc_0$ with $90$\% prior probability. If one wishes to place such probability mass on partitions within distance $1$ from $\boldc_0$, a value of $\psi$ around $10$ and $15$ is needed, respectively, under uniform and DP base prior. We suggest, when performing the analysis, to consider also neighborhood values of the chosen $\psi$, in order to assess the sensitivity of the results.

\section{The National Birth Defects Prevention Study}\label{sec5:nbdps}

The National Birth Defects Prevention Study (NBDPS) is a multi-state population-based, case-control study of birth defects in the United States \citep{yoon2001}. Infants were identified using birth defects surveillance systems in recruitment areas within ten US states (Arkansas, California, Georgia, Iowa, Massachusetts, New Jersey, New York, North Carolina, Texas, and Utah), which cover roughly 10\% of US births. Diagnostic case information was obtained from medical records and verified by a standardized clinician review specific to the study \citep{rasmussen2003}. Participants in the study included mothers with expected dates of delivery from 1997-2009. Controls were identified from birth certificates or hospital records and were live-born infants without any known birth defects.
Each state site attempted to recruit $300$ cases and $100$ (unmatched) controls annually.
A telephone interview was conducted with case and control mothers to solicit a wide range of demographic, lifestyle, medical, nutrition, occupational and environmental exposure history information.

Because birth defects are highly heterogeneous, a relatively large number of defects of unknown etiology are included in the NBDPS.
We are particularly interested in congenital heart defects (CHD), the most common type of birth defect and the leading cause of infant death due to birth defects.
Because some of these defects are relatively rare, in many cases we lack precision for investigating associations between potential risk
factors and individual birth defects. For this reason, researchers
typically lump embryologically distinct and potentially etiologically heterogeneous defects in order to increase power (e.g., grouping all heart defects together), even knowing the underlying mechanisms may differ substantially. In fact, how best to group defects is subject to uncertainty, despite a variety of proposed groupings available in the literature \citep{lin1999cardiovascular}.

In this particular application, we consider $26$ individual heart defects, which have been previously grouped into $6$ categories by investigators \citep{botto2007}. The prior grouping is shown in Table~\ref{tab:defects}, along with basic summary statistics of the distribution of defects in the analyzed data.
We are interested in evaluating the association between heart defects and about $90$ potential risk factors related to mothers' health status, pregnancy experience, lifestyle and family history.
We considered a subset of data from NBDPS, excluding observations with missing covariates, obtaining a dataset with $8,125$ controls, while all heart defects together comprise $4,947$ cases.

\begin{center}
 \begin{table}[t]
  \resizebox{\columnwidth}{!}{
   \begin{tabular}{@{}lccc@{}}
    \toprule
    \multicolumn{1}{l}{Congenital Heart Defect \hspace*{0.30em}} & {Abbreviation} & Frequency & Percentage of cases \\
    \midrule
    \multicolumn{1}{l}{\textbf{Septal}}                                                                               \\
    Atrial septal defect                                         & ASD           & 765         & 0.15                \\
    Perimembranous ventricular septal defect                     & VSDPM          & 552         & 0.11                \\
    Atrial septal defect, type not specified                     & ASDNOS         & 225         & 0.04                \\
    Muscular ventricular septal defect                           & VSDMUSC        & 68          & 0.02                \\
    Ventricular septal defect, otherwise specified               & VSDOS          & 12          & 0.00                \\
    Ventricular septal defect, type not specified                & VSDNOS         & 8           & 0.00                \\
    Atrial septal defect, otherwise specified                    & ASDOS          & 4           & 0.00                \\
    \midrule
    \multicolumn{1}{l}{\textbf{Conotruncal}}                                                                          \\
    Tetralogy of Fallot                                          & FALLOT         & 639         & 0.12                \\
    D-transposition of the great arteries                        & DTGA           & 406         & 0.08                \\
    Truncus arteriosus                                           & COMMONTRUNCUS  & 61          & 0.01                \\
    Double outlet right ventricle                                & DORVTGA        & 35          & 0.01                \\
    Ventricular septal defect reported as conoventricular        & VSDCONOV       & 32          & 0.01                \\
    D-transposition of the great arteries, other type            & DORVOTHER      & 22          & 0.00                \\
    Interrupted aortic arch type B                               & IAATYPEB       & 13          & 0.00                \\
    Interrupted aortic arch, not otherwise specified             & IAANOS         & 5           & 0.00                \\
    \midrule
    \multicolumn{1}{l}{\textbf{Left ventricular outflow}}                                                             \\
    Hypoplastic left heart syndrome                              & HLHS           & 389         & 0.08                \\
    Coarctation of the aorta                                     & COARCT         & 358         & 0.07                \\
    Aortic stenosis                                              & AORTICSTENOSIS & 224         & 0.04                \\
    Interrupted aortic arch type A                               & IAATYPEA       & 12          & 0.00                \\
    \midrule
    \multicolumn{1}{l}{\textbf{Right ventricular outflow}}                                                            \\
    Pulmonary valve stenosis                                     & PVS            & 678         & 0.13                \\
    Pulmonary atresia                                            & PULMATRESIA    & 100         & 0.02                \\
    Ebstein anomaly                                              & EBSTEIN        & 66          & 0.01                \\
    Tricuspid atresia                                            & TRIATRESIA     & 46          & 0.01                \\
    \midrule
    \multicolumn{1}{l}{\textbf{Anomalous pulmonary venous return}}                                                    \\
    Total anomalous pulmonary venous return                      & TAPVR          & 163         & 0.03                \\
    Partial anomalous pulmonary venous return                    & PAPVR          & 21          & 0.01                \\
    \midrule
    \multicolumn{1}{l}{\textbf{Atrioventricular septal defect}}                                                       \\
    Atrioventricular septal defect                               & AVSD           & 112         & 0.02                \\
    \bottomrule
   \end{tabular}
  }
  \caption{Summary statistics of the distribution of congenital heart defects among cases. Defects are divided according the grouping provided from investigators.}
  \label{tab:defects}
 \end{table}
\end{center}

\subsection{Modeling birth defects}

Standard approaches assessing the impact of exposure factors on the risk to develop a birth defect often rely on logistic regression analysis. Let $i = 1, \ldots, N$ index birth defects, while $j = 1, \ldots, n_i$ indicates observations related to birth defect $i$, with $y_{ij} = 1$ if observation $j$ has birth defect $i$ and $y_{ij} = 0$ if observation $j$ is a control, i.e. does not have any birth defect. Let $\mathbf{X}_i$ denote the data matrix associated to defect $i$, with each row $\mathbf{x}^T_{ij} = (x_{ij1}, \ldots, x_{ijp})$ being the vector of the observed values of $p$ categorical variables for the $j$th observation.
At first one may consider $N$ separate logistic regressions of the type
\begin{equation}\label{eq:glm}
 \log\left(\frac{\Pr(y_{ij} = 1|\mathbf{x}_{ij})}{\Pr(y_{ij} = 0|\mathbf{x}_{ij})} \right) = \text{logit}(\boldsymbol{\pi}_{ij}) = \alpha_i + \mathbf{x}_{ij}^{T} \boldsymbol{\beta}_i,
\end{equation}
with $\alpha_i$ denoting the defect-specific intercept, and $\boldsymbol{\beta}_i$ the $p\times 1$ vector of regression coefficients. However, Table~\ref{tab:defects} highlights the heterogeneity of heart defect prevalences, with some of them being so few as to preclude separate analyses.

A first step in introducing uncertainty about clustering of the defects may rely on a standard Bayesian nonparametric approach, placing a Dirichlet process prior on the distribution of regression coefficient vector $\boldsymbol{\beta}_i$ in order to borrow information across multiple defects while letting the data inform on the number and composition of the clusters. A similar approach has been previously proposed in \cite{maclehose2010}, with the aim being to shrink the coefficient estimates towards multiple unknown means. In our setting, an informed guess on the group structure is available through $\boldc_0$, reported in Table~\ref{tab:defects}.

We consider a simple approach building on the Bayesian version of the model in \eqref{eq:glm}, and allowing the exposure coefficients $\boldsymbol{\beta}_i$ for $i = 1, \ldots, N$ to be shared across regressions, while accounting for $\boldc_0$. The model written in a hierarchical form is
\begin{alignat}{3}\label{eq:dp_logistic}
 y_{ij}    & \sim Ber(\pi_{ij}) \quad                 & \text{logit}(\pi_{ij}) & =\alpha_i + \mathbf{x}_{ij}^{T} \boldsymbol{\beta}_{c_i}, \quad j = 1, \ldots, n_i, \nonumber \\
 \alpha_i  & \sim \mathcal{N}(a_0, \tau_0^{-1}) \quad & \boldsymbol{\beta}_{c_i} |  \boldc &\sim \mathcal{N}_p (\mathbf{b}, \mathbf{Q})  \quad i = 1, \ldots, N,     \nonumber                 \\
 p(\boldc)  & \sim CP(\boldc_0, \psi, p_0(\boldc))  \quad & p_0(\boldc) &\propto \alpha^{K} \prod_{k = 1}^{K} (\lambda_k - 1)!
\end{alignat}

where $CP(\boldc_0, \psi, p_0(\boldc_0))$ indicates the Centered Partition process, with base partition $\boldc_0$, tuning parameter $\psi$ and baseline EPPF $p_0(\boldc_0)$. We specify the baseline EPPF so that when $\psi = 0$ the prior distribution reduces to a Dirichlet Process with concentration parameter $\alpha$. Instead, for $\psi \rightarrow \infty$ the model corresponds to $K$ separate logistic regressions, one for each group composing $\boldc_0$.
The model estimation can be performed by leveraging a P{\`o}lya-Gamma data-augmentation strategy for Bayesian logistic regression \citep{polson2013}, combined with the procedure illustrated in Section~\ref{sec4:Posterior computation} for the clustering update step. The Gibbs sampler is detailed in the Appendix (Algorithm~\ref{alg:gibbs}).

\subsection{Simulation study}\label{sec:simu}

We conduct a simulation study to evaluate the performance of our approach in accurately estimating the impact of the covariates across regressions with common effects, under different prior guesses.
In simulating data, we choose a scenario mimicking the structure of our application, considering a number of defects $N = 12$ equally partitioned in $4$ groups.
We consider $p = 10$ dichotomous explanatory variables and assume that defects in the same group have the same covariates effects.
We take a different number of observations across defects, with $\{n_1, n_2, n_3\} = \{100, 600, 200\}$, $\{n_4, n_5, n_6\}  = \{300, 100, 100\}$, $\{n_7, n_8, n_9\} = \{500, 100, 200\}$, $\{n_{10}, n_{11}, n_{12}\} = \{200, 200, 200\}$.
For each defect $i$ with $i = 1, \ldots, 12$ we generate a data matrix $\mathbf{X}_i$  by sampling each of the variables from a Bernoulli distribution with probability of success equal to $0.5$. We set most of coefficients $\beta_{i1}, \ldots, \beta_{i10}$ to $0$, while defining a challenging scenario with small to moderate changes across different groups.
In particular we fix $\{\beta_1, \beta_2, \beta_3, \beta_4\} = \{ 0.7, -1.2, 0.5, 0.5 \}$ for group $1$, $\{\beta_4, \beta_5, \beta_6\} = \{ 0.7, -0.7, 0.7\}$ for group $2$, $\{\beta_9, \beta_{10}\} = \{0.7, -1.2\}$ for group $3$ and $\{\beta_1, \beta_2, \beta_9,  \beta_{10}\} = \{ 0.7, -0.7, 0.7, -0.7\}$ for group $4$.
Finally response variables $\mathbf{y}_i$ for $i = 1, \ldots, 12$ are drawn from a Bernoulli distribution with probability of success $p_i = \text{logit}(\mathbf{X}_i^T \boldsymbol{\beta}_i)$.

We compare coefficients and partition estimates from a grouped logistic regression using a DP prior with $\alpha = 1$ and using a CP prior with DP base EPPF with $\alpha = 1$. In evaluating the CP prior performances, we consider both the true known partition and a wrong guess. Posterior estimates are obtained using the Gibbs sampler described in the Appendix. We consider a multivariate normal distribution with zero mean vector and covariance matrix $\mathbf{Q} = \text{diag}_p(2)$ as base measure for the DP, while we assume the defect-specific intercepts $\alpha_i \sim N(0, 2)$ for $i = 1, \ldots, 12$.
We run the algorithm for $5,000$ iterations discarding the first $1,000$ as burn-in, with inspection of trace-plots suggesting convergence of the parameters.

\begin{figure}[H]
 \includegraphics[width = 1\textwidth]{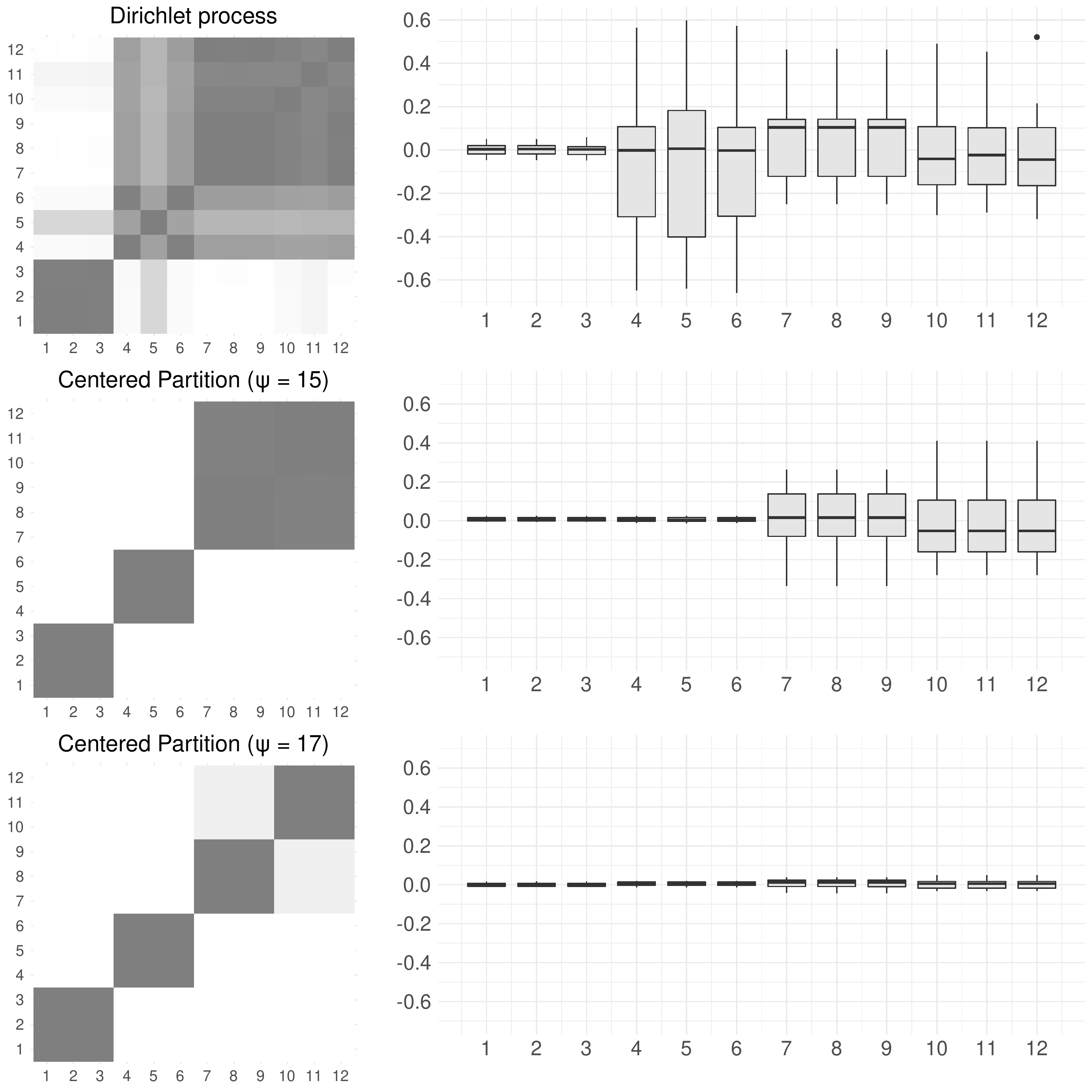}
 \caption{Results from grouped logistic regressions with DP($\alpha = 1$) prior with and CP process prior with DP($\alpha = 1$) base EPPF for $\psi = \{15, 17\}$, centered on the true partition. Heatmaps on the left side show the posterior similarity matrix. On the right side, boxplots show the distribution of deviations from the maximum likelihood baseline coefficients and posterior mean estimates for each defect $i = 1, \ldots, 12$.}\label{fig:sim_cp}
\end{figure}

In evaluating the resulting estimates under different settings, we take as baseline values for coefficients the maximum likelihood estimates obtained under the true grouping. Figure~\ref{fig:sim_cp} shows the posterior similarity matrices obtained under the Dirichlet and Centered Partition processes, along with boxplots of the distribution of differences between the coefficients posterior mean estimates and their baseline values, for each of the $12$ simulated defects.
We first centered the CP prior on the true known grouping and, according to the considerations made in Section~\ref{sec:prior_monte_carlo}, we fixed the value of $\psi$ to $15$ for the CP process prior, founding the maximum a posteriori estimate of the partition almost recovering the true underlying grouping expect for merging together the third and fourth group.

We also considered other values for $\psi$ close to $15$, and report the case for $\psi = 17$ in Figure~\ref{fig:sim_cp}, for which the true grouping is recovered, with resulting mean posterior estimates of the coefficients almost identical to the baseline. When considering the Dirichlet process, although borrowing information across the defects, it does not distinguish between all the groups but individuate only the first one, while the CP process recovers the true grouping, with better performances in estimating the coefficients.

Finally, we evaluate the CP prior performances when centered on a wrong guess $\boldc_0^\prime$ of the base partition. In particular, we set $\boldc_0^\prime = \{1,5,9\}\{2,6,10\}\{3,7,11\}\{4,8,12\}$. Despite having the same configuration of $\boldc_0$, it has distance from $\boldc_0$ of approximately $3.16$, where the maximum possible distance is $\log_2(12) = 4.70$.
Under such setting we estimate the partition $\hat{\boldc} = \{1,2,3,5\}\{4,6,7,8,9,10,11,12\}$ via maximum at posteriori, obtaining two clusters.
Although we center the prior in $\boldc_0^\prime$, the estimated partition results to be closer to the one induced by the DP ($0.65$) than $\boldc_0^\prime$ ($2.45$), with also similar performances in the coefficient estimation, which may be interpreted as a suggestion that the chosen base partition is not supported by the data.

\begin{figure}[!tb]
 \includegraphics[width = 1\textwidth]{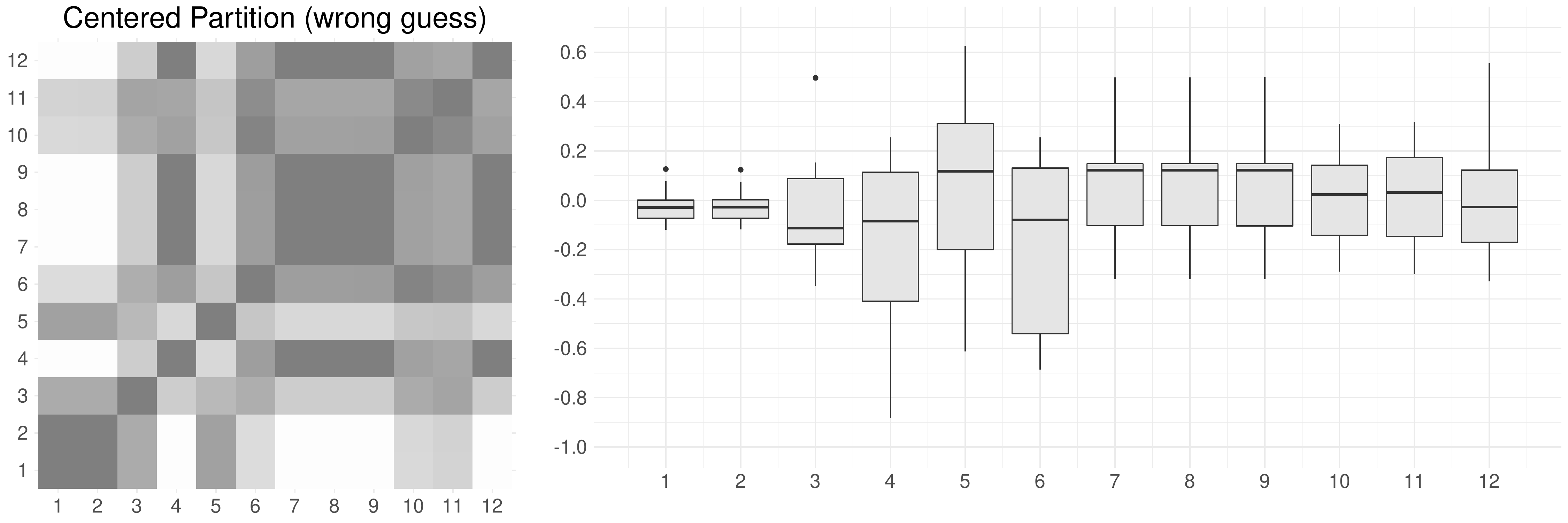}
 \caption{
Results from grouped regression using CP process prior with DP($\alpha = 1$) base EPPF for $\psi = 15$ centered on partition $\boldc_0^\prime = \{1,5,9\}\{2,6,10\}\{3,7,11\}\{4,8,12\}$ which has distance $3.16$ from the true one. Heatmaps on the left side show the posterior similarity matrix. On the right side, boxplots show the distribution of deviations from the maximum likelihood baseline coefficients and posterior mean estimates for each defect $i = 1, \ldots, 12$.}\label{fig:sim_cp_wrong}
\end{figure}

\subsection{Application to NBDPS data}

We estimated the model in \eqref{eq:dp_logistic} on the NBDPS data, considering the controls as shared with the aim of grouping cases into informed groups on the basis of the available $\boldc_0$. In order to choose a value for the penalization parameter, we consider the prior calibration illustrated in Section~\ref{sec:prior_calibration}, finding a value of $\psi = 40$ assigning a $90$\% probability to partitions within a distance around $0.8$, where the maximum possible distance is equal to $4.70$. In terms of moves on the Hasse diagram we are assigning $90\%$ prior probability to partitions at most at $11$ split/merge operations from $\boldc_0$, given that the minimum distance from $\boldc_0$ is $2/N \approx 0.07$.
To assess sensitivity of the results, we performed the analysis under different values of $\psi \in \{0, 40, 80, 120, \infty\}$. In particular, for $\psi = 0$ the clustering behavior is governed by a Dirichlet process prior, while $\psi \rightarrow \infty$ corresponds to fixing the groups to $\boldc_0$.

\begin{figure}[H]
 \captionsetup[sub]{font=small,labelfont={bf,sf}}
 \subfloat[$\psi$ = 0, $\text{VI}(\hat{\boldc}, \boldc_0) = 2.43$]{\includegraphics[width = 0.48\textwidth]{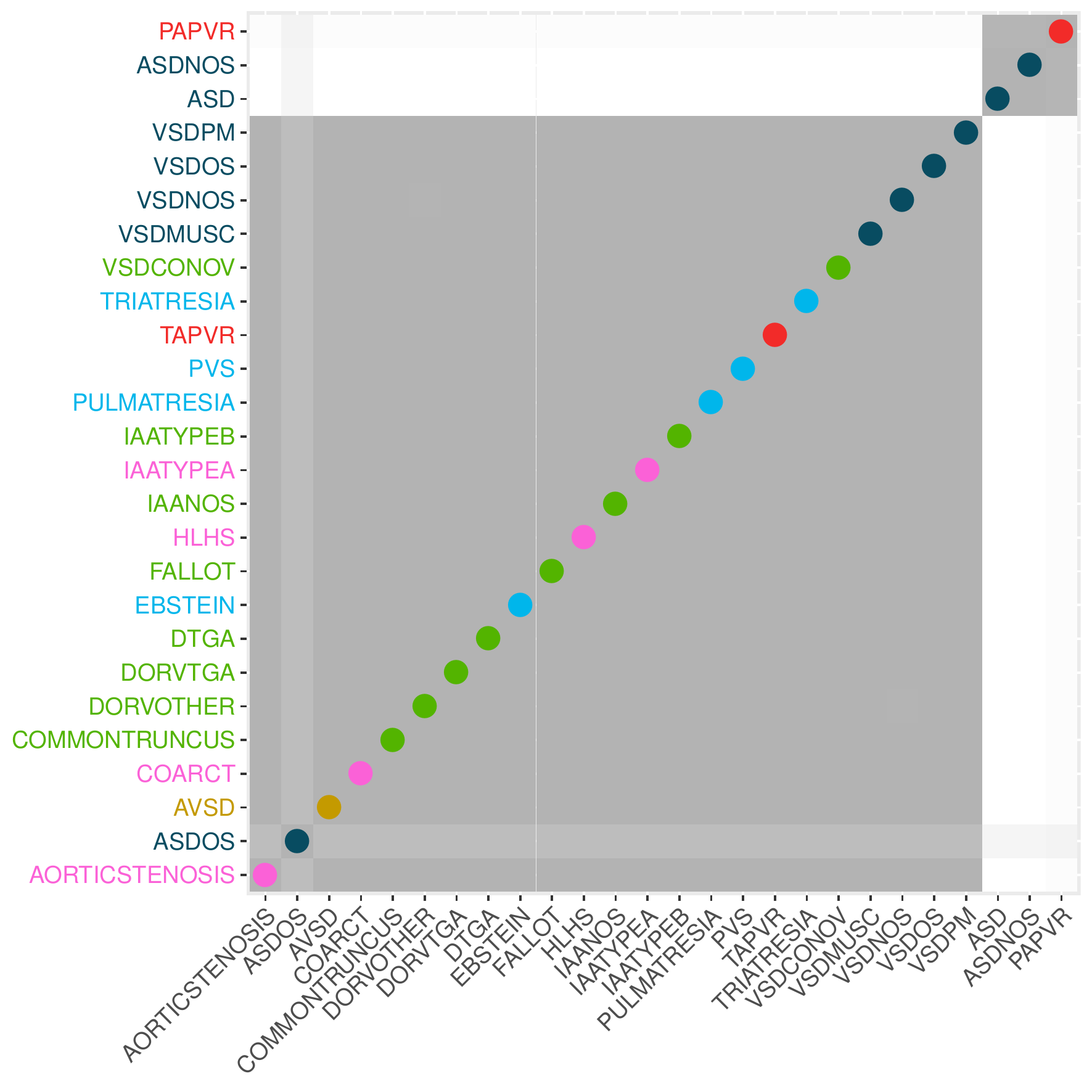}}
 \subfloat[$\psi$ = 40, $\text{VI}(\hat{\boldc}, \boldc_0) = 1.78$]{\includegraphics[width = 0.48\textwidth]{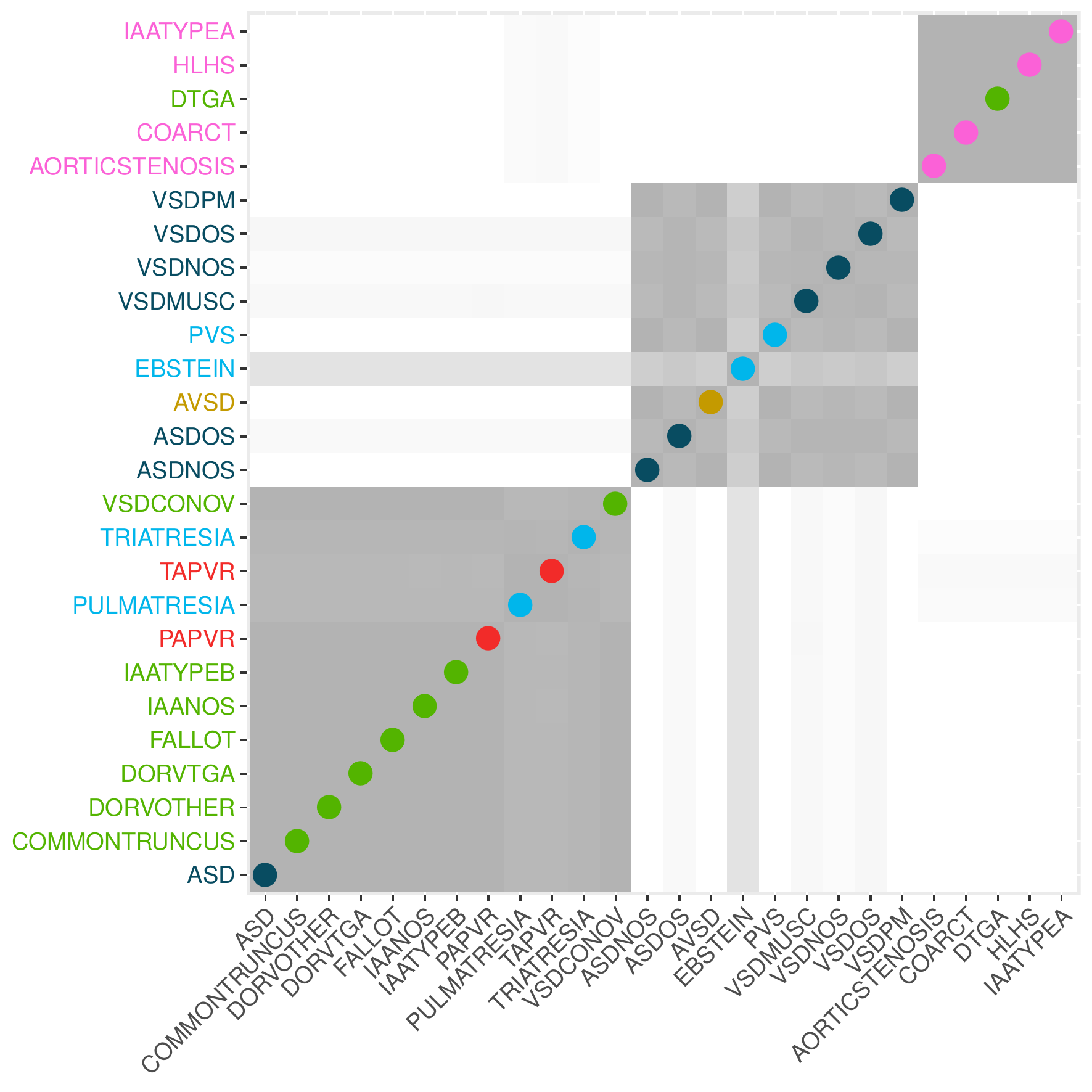}}\\
 \subfloat[$\psi$ = 80, $\text{VI}(\hat{\boldc}, \boldc_0) = 1.65$]{\includegraphics[width = 0.48\textwidth]{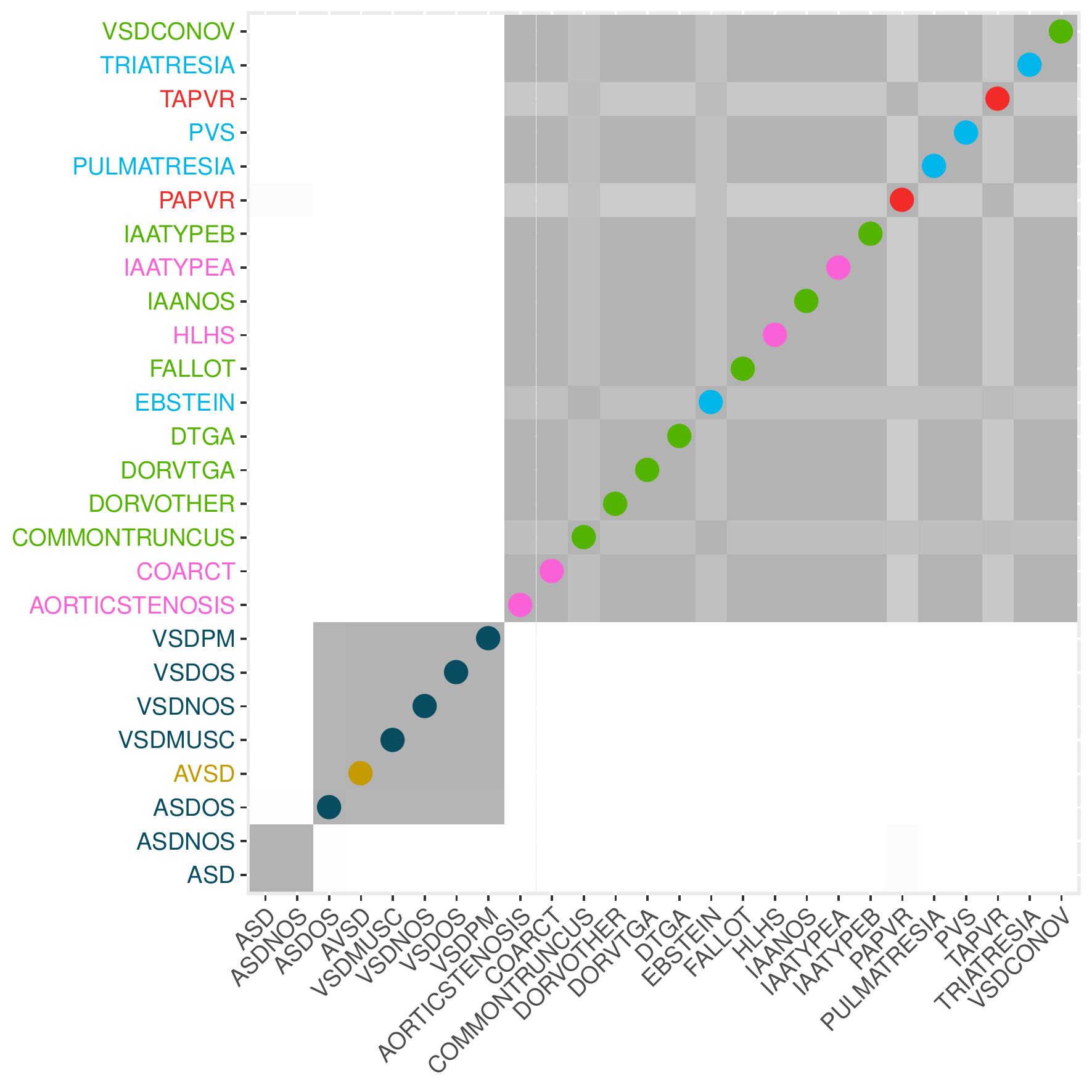}}
 \subfloat[$\psi$ = 120, $\text{VI}(\hat{\boldc}, \boldc_0) = 0.86$]{\includegraphics[width = 0.48\textwidth]{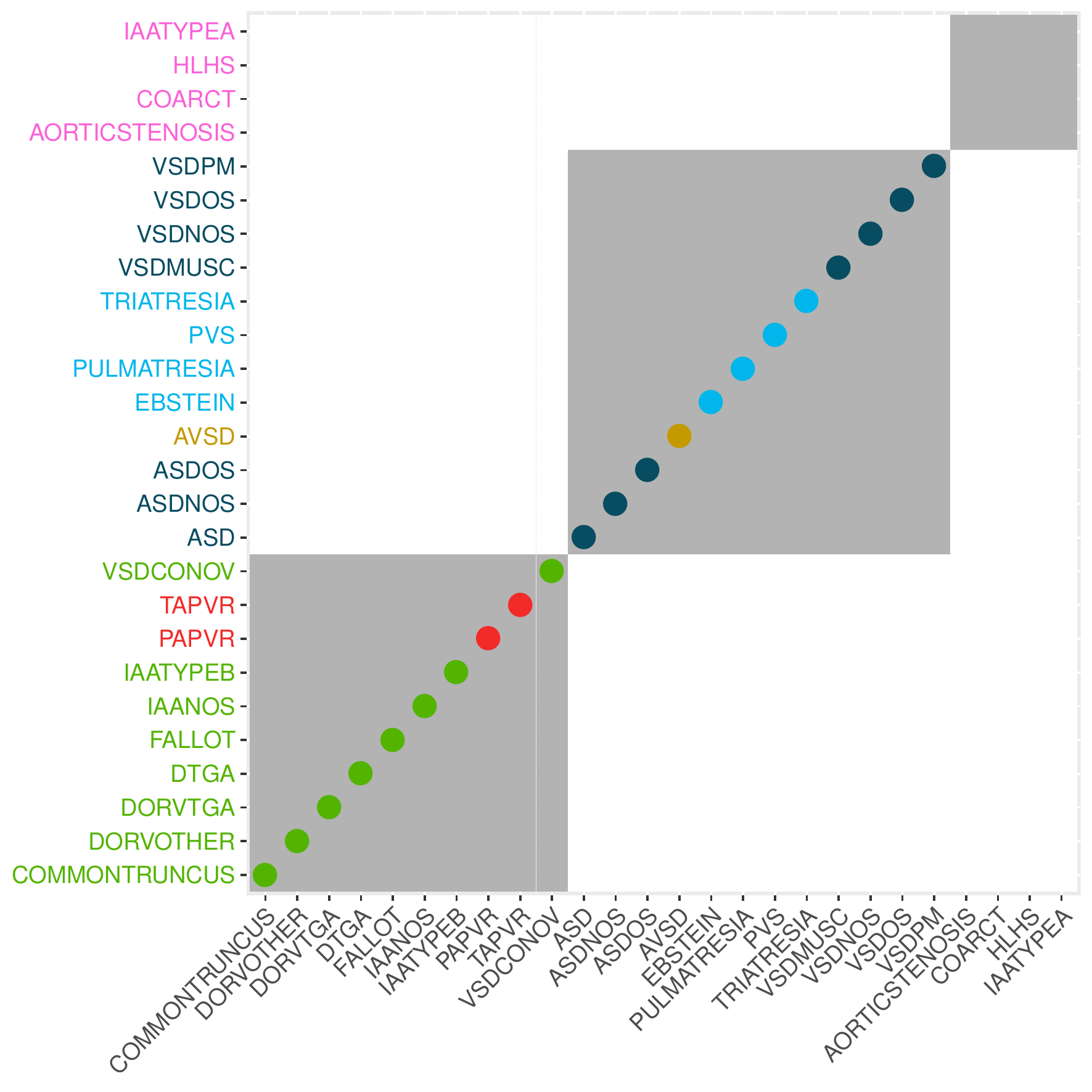}}
 \caption{Posterior allocation matrices obtained using the CP process with a DP ($\alpha = 1$) prior for different values of $\psi \in \{0, 40, 80, 120\}$. On the y-axis labels are colored according base grouping information $\boldc_0$, with dots on the diagonal highlighting differences between $\boldc_0$ and the estimated partition $\hat{\boldc}$.}\label{fig:cp_clustering}
\end{figure}

In analyzing the data we run the Gibbs sampler for $10,000$ iterations and use a burn-in of $4,000$, under the same prior settings as in Section~\ref{sec:simu}. Figure~\ref{fig:cp_clustering} summarizes the posterior estimates of the allocation matrices under different values of $\psi$, with colored dots emphasizing differences with the base partition $\boldc_0$. Under the DP process ($\psi = 0$) the estimated partition differs substantially from the given prior clustering.
Due to the immense space of the possible clusterings, this is likely reflective of limited information in the data, combined with the tendency of the DP to strongly favor certain types of partitions, typically characterized from few large clusters along many small ones. When increasing the value of the tuning parameter $\psi$ the estimated clustering is closer to $\boldc_0$, with a tendency in favoring a total number of three clusters. In particular, for $\psi = 120$ one of the groups in $\boldc_0$ is recovered (left ventricular outflow), while the others are merged in two different groups. It is worth noticing that AVSD, which is placed in its own group under $\boldc_0$, is always grouped with other defects with a preference for ones in the septal class (blue color). Also two defects of this last class, ASD and ASDOS, happen to be lumped together across different values of $\psi$, and are in fact two closely related defects.

\begin{figure}
 \includegraphics[width = \textwidth]{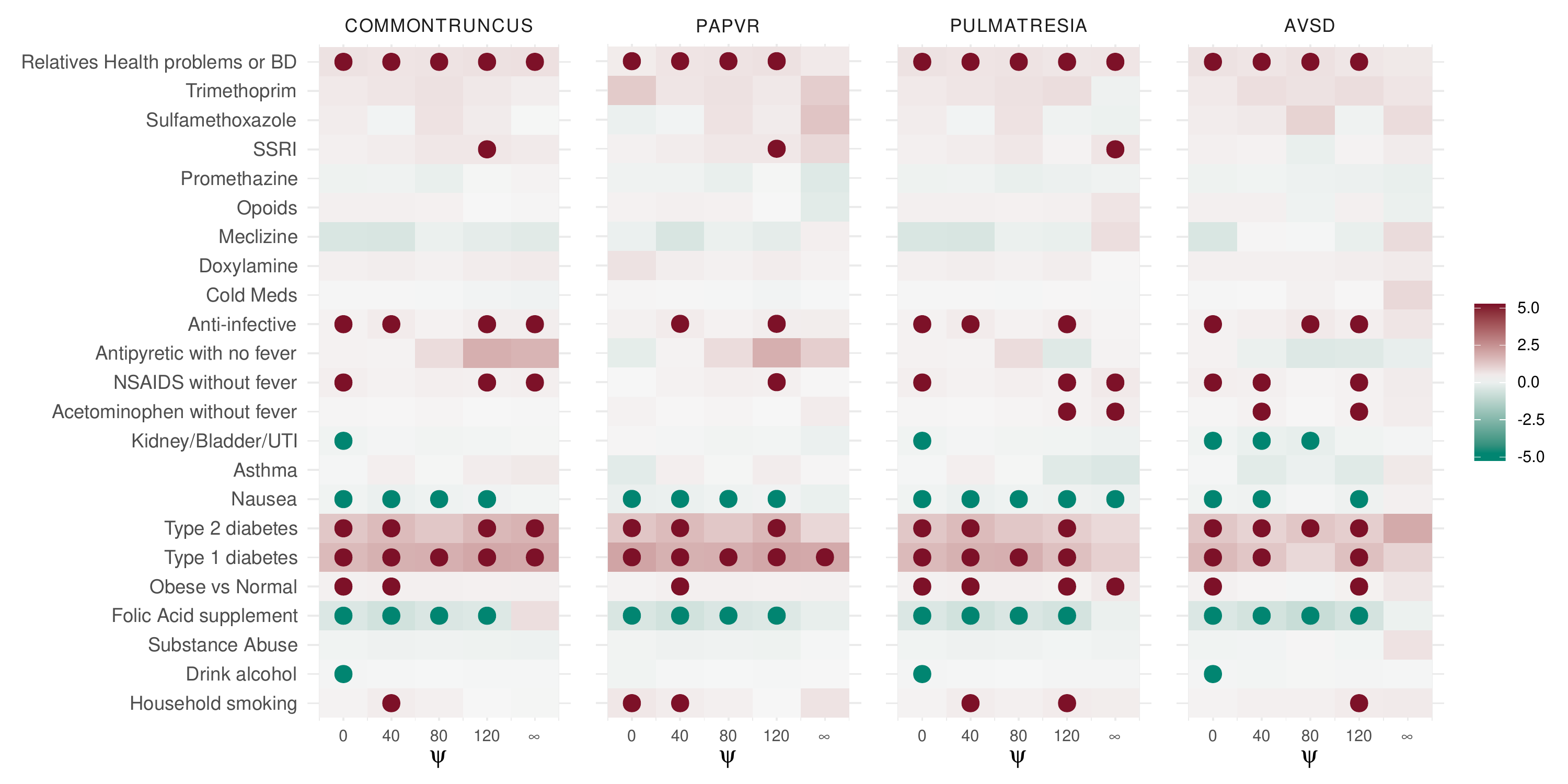}
 \caption{Comparison of significant odds ratio under $\psi \in\{0, 40, 80, 120,\infty\}$ for some exposure factors and $4$ selected heart defects in $4$ different groups under $\boldc_0$. Dots are in correspondence of significant mean posterior log-odds ratios (log-OR) at $95\%$ with red encoding risk factors (log-OR $> 0$) and green protective factors (log-OR $< 0$).}\label{fig:odds_trend}
\end{figure}

Details on the results for each of the estimated models are given in the Appendix (Figures~\ref{fig:odds_dp}-\ref{fig:pg}) and summarized here. Figure~\ref{fig:odds_trend} shows a heatmap of the mean posterior log odds-ratios for increasing values of the penalization parameter $\psi$, with dots indicating if they are significant according to a $95\%$ credibility interval. In general, the sign of the effects does not change for most of the exposure factors across the different clusterings. Figure~\ref{fig:odds_trend} focuses on pharmaceutical use in the period from $1$ month before the pregnancy and $3$ months during, along with some exposures related to maternal behavior and health status.

We found consistent results for known risk factors for CHD in general, including for diabetes \citep{correa2008diabetes} and obesity \citep{waller2007prepregnancy}. The finding that nausea is associated with positive outcomes is consistent with prior literature \citep{koren}. The association between use of SSRIs and pulmonary atresia was also noted in \cite{reefhuis2015specific}. It is worth noticing that estimates obtained under the DP prior are less consistent with prior work. In particular, there apparent artifacts such as the protective effect of alcohol consumption related to defects in the bigger cluster, which is mitigated from an informed borrowing across the defects.
On the other side, estimates under separate models for AVSD or PAPVR , which corresponds to $0.02\%$ and $0.01\%$ of cases respectively, show how a separate analysis of cases with low prevalence misses even widely assessed risk factors, as for example diabetes.

\section*{Discussion}

There is a very rich literature on priors for clustering, with almost all of the emphasis on exchangeable approaches, with a smaller literature focused on including dependence on known features (eg., temporal or spatial structure or covariates). The main contribution of this article is to propose what is seemingly a first attempt at including prior information on an informed guess at the clustering structure. We were particularly motivated by a concrete application to a birth defects study in proposing our method, which is based on shrinking an initial clustering prior towards the prior guess.

There are many immediate interesting directions for future research. One thread pertains to developing better theoretical insight and analytical tractability into the new class of priors. For existing approaches, such as product partition models and Gibbs-type partitions, there is a substantial literature providing simple forms of prediction rules and other properties. It is an open question whether such properties can be modified to our new class. This may yield additional insight into the relative roles of the base prior, centering value and hyperparameters in controlling the behavior of the prior and its impact on the posterior.

Another important thread relates to applications of the proposed framework beyond the setting in which we have an exact guess at the complete clustering structure. In many cases, we may have an informed guess or initial clustering in a subset of the objects under study, with the remaining objects (including future ones) completely unknown. Conceptually the proposed approach can be used directly in such cases, and also when one has different types of prior information on the clustering structure than simply which objects are clustered together.

\subsection*{acknowledgement}
 The authors gratefully acknowledge this work was supported in part through cooperative agreements from the Centers for Disease Control and Prevention to the centers participating in the National Birth Defects Prevention Study and by the National Institutes of Health (R01ES027498; U50CCU422096; 5U01DD001036; PA96043; PA 02081; FOA DD09-001).

\bibliographystyle{ba}
\bibliography{references}

\newpage
\section*{Appendix}

\subsection*{Prior calibration}

\begin{breakablealgorithm}
 \caption{: Estimation of counts statistics related to distances neighborhoods of $\boldc_0$}
 \begin{algorithmic}\label{alg:prior_est}
   \STATE \textbf{Local search}
  \STATE 0. Start from the base partition $\boldc_0$ with $|K_0|$ clusters and configuration $\boldsymbol{\lambda}_{m_0}$ and set $\delta_0 = 0$ and $\mathcal{N}_0(\boldc_0) = \boldc_0$.
  \FOR{$t = 1, \ldots, T$}
  \STATE Obtain $\mathcal{N}_t(\boldc_0)$ from partitions in $\mathcal{N}_{t-1}(\boldc_0)$ by exploring all directed connections, i.e. partitions obtained with one operation of split/merge on elements $\mathcal{N}_{t-1}(\boldc_0)$.
  \ENDFOR
  \STATE 2. Compute the distance from $\boldc_0$ and all partitions in $\mathcal{N}_T(\boldc_0)$ and take the minimum distance, $\delta_{L^*}$; discard all partitions having distances greater than $\delta_{L^*}$.
  \STATE 3. Obtain counts $n_{l}$ and $n_{lm}$ relative to distances $\delta_1, \ldots, \delta_{L^*}$ for $m = 1, \ldots, M$.
  \STATE \textbf{Monte Carlo approximation}
  \FOR{$r = 1, \ldots, R$}
  \STATE 4. Sample the number of clusters $K$ from the discrete probability distribution  $$p(K = k) = e^{-1}k^{N}/(k!\mathcal{B}_N),\quad k \in \{1, \ldots, N\}.$$
  \vspace{-4ex}
  \STATE 5. Conditional on $K$ generate a partition $\boldc^{(r)} = \{c_1^{(r)}, \ldots, c_N^{(r)}\}$ by sampling each $c^{(r)}_i$ from a discrete uniform distribution on $\{1, \ldots, K\}$.
  \STATE 6. If $d(\boldc^{(r)}, \boldc_0) > \delta_{L^*}$ reject the partition.
  \ENDFOR
  \STATE 7. Let $R^*$ be the number of accepted partitions, and estimate counts $\hat{n}_{l}$ and $\hat{n}_{lm}$ for $m = 1, \ldots, M$ and according to~\eqref{eq:est_nl}-\eqref{eq:est_nlm} conditional on the observed distance values $\hat{\delta}_{(L^* + 1)}, \ldots, \hat{\delta}_L$.
  \STATE 8. Using $R^*$ be the number of accepted partitions, and estimate counts $\hat{n}_{l}$ and $\hat{n}_{lm}$ relative to distances $\hat{\delta}_{L^* + 1}, \ldots, \hat{\delta}_L$ for $m = 1, \ldots, M$.
 \end{algorithmic}
\end{breakablealgorithm}

\subsection*{Marginal sampling using variation of information}

We describe how to compute the penalization term in the marginal sampling step described in Section~\ref{sec4:Posterior computation} using the Variation of Information as a distance, but the same procedure applies when using other distances based on blocks sizes.
Let $K^{-}$ and $K_0^{-}$ denote respectively the number of clusters in $\boldc^{-i}$ and $\boldc_0^{-i}$, i.e. partitions $\boldc$ and $\boldc_0$ after removing the $i$ observation.

\begin{breakablealgorithm}
 \caption{: Computation strategy for the penalization term in marginal sampling}
 \begin{algorithmic}\label{alg:distance}
 \STATE Let $K^{-}$ and $K_0^{-}$ denote respectively the number of clusters in $\boldc^{-i}$ and $\boldc_0^{-i}$, i.e. partitions $\boldc$ and $\boldc_0$ after removing the $i$ observation.
  \FOR{$i = 1, \ldots, N$}
  \STATE 1. Compute cardinalities $\{\lambda_1^{-i}, \ldots, \lambda_{K^{-}}^{-i}\}$ representing the number of observations in each cluster for $\boldc^{-i}$.
  \STATE 2. Compute $\lambda_{lm}^{-i}$, the number of observations in cluster $l$ under $\boldc^{-i}$ and cluster $m$ under $\boldc_0^{-i}$ for $l = 1, \ldots, K^{-}$ and $m = 1, \ldots, K_0^{-}$.
  \FOR{$k = 1, \ldots, K^{-}, K^{-} + 1$}
  \STATE Let $c_{i,0}$ be the cluster of index $i$ under partition $\boldc_0$.\\
  Compute $d(\boldc, \boldc_0) \propto -H(\boldc) +2H(\boldc \land \boldc_0)$ for $\boldc= \{\boldc^{-i} \cup k\}$ using
  \begin{align*}
   - H(\boldc)              = & \sum_{l \neq k}^K \left\{\frac{\lambda_l^{-i}}{N} \log \frac{\lambda_l^{-i}}{N} \right\}+ \left(\frac{\lambda_k^{-i} + 1}{N}  \right) \log\left(\frac{\lambda_k^{-i} + 1}{N}  \right) \\
   H(\boldc \land \boldc_0) = & - \Bigg\{ \sum_{l = 1}^{K} \sum_{m =1}^{K_0^{-}} \frac{\lambda_{lm}^{-i}}{N} \log \left(\frac{\lambda_{lm}^{-i}}{N}\right) -
   \frac{\lambda_{kc_{i,0}}^{-i}}{N} \log \left(\frac{\lambda_{kc_{i,0}}^{-i}}{N}\right)                                                                                                                  \\
                              & + \frac{\lambda_{kc_{i,0}}^{-i} + 1}{N} \log \left(\frac{\lambda_{kc_{i,0}}^{-i} + 1}{N}\right)
   \Bigg\}
  \end{align*}
  \ENDFOR
  \ENDFOR
 \end{algorithmic}
\end{breakablealgorithm}
\subsection*{Gibbs sampling for shared logistic regression}
In estimating the model, a P{\'o}lya-gamma data augmentation strategy is employed; for each $y_{ij}$ we introduce a latent variable $\omega_{ij} \sim PG(1, \alpha_i + \mathbf{x}_{ij}^{T} \boldsymbol{\beta}_{c_i})$ for each observation $j$ in defect-specific dataset $i$ for $i = 1, \ldots, N$.
\\
\begin{breakablealgorithm}\label{alg:gibbs}
 \caption{: Gibbs sampling for posterior computation}
 \begin{algorithmic}
  \STATE Conditionally on the cluster allocation vector $\mathbf{c} = (c_1, \ldots, c_n)$ and data $\{ \mathbf{y}_i, \mathbf{X}_i\}$ for $i = 1, \ldots, N$, update mixture related parameters and P{\'o}lya-gamma latent variables as follows.
  \STATE --------------------------------------------------------------------------------------------------------
  \STATE { [1] Sample P{\'o}lya-gamma latent variables for each observation in each dataset}
  \FOR{$i = 1, \ldots, N$ and $j = 1, \ldots, n_i$}
  \STATE{
   \begin{equation*}
    (\omega_{ij}| - ) \sim PG(1, \alpha_i + \mathbf{x}_{ij}^{T} \boldsymbol{\beta}_{c_i})
   \end{equation*}
  }
  \ENDFOR
  \STATE --------------------------------------------------------------------------------------------------------
  \STATE { [2] Update defect-specific intercept, exploiting P{\'o}lya-gamma conjugancy}
  \FOR{$i = 1, \ldots, N$}
  \STATE
  {
  \begin{equation*}
   (\alpha_i| - ) \sim \mathcal{N}(a^*, \tau^*)
  \end{equation*}
  with $\tau^* = \tau_0 + \sum_{j = 1}^{n_i} \omega_{ij}$ and $a^* = [a_0 \tau_0 + \sum_{j = 1}^{n_i} (y_{ij} - 1/2 - \omega_{ij} \mathbf{x}_{ij}^{T}
   \boldsymbol{\beta}_{c_i} )]/ \tau^*$
  }
  \ENDFOR
  \STATE --------------------------------------------------------------------------------------------------------
  \STATE { [3] Defining $\kappa_{ij} := y_{ij} - 1/2 - w_{ij} \alpha_i$, then the vector ($\kappa_{ij}/\omega_{ij}| c_i = h, \omega_{ij}) \sim \mathcal{N}(\mathbf{x}_i^{T}\boldsymbol{\beta}^{(k)}, 1/\omega_{ij})$, and each cluster-specific coefficient vector $\boldsymbol{\beta}_h$
   can be updated by aggregating all observations and augmented data relative to birth defects that are in the same cluster.}
  \FOR{$k = 1, \ldots, K$}
  \STATE{Let $\mathbf{X}^{(k)}$, $\mathbf{y}^{(k)}$, $\boldsymbol{\kappa}^{(k)}$ be the obtained quantities relative to cluster $h$, and $\mathbf{\Omega}^{(k)}$ a diagonal matrix with the corresponding P{\'o}lya-gamma augmented variables.
  Then update cluster-specific coefficients vector from
  \begin{equation*}
   (\boldsymbol{\beta} ^{(k)}| - ) \sim \mathcal{N}_p (\mathbf{b}^{(k)}, \mathbf{Q}^{(k)})
  \end{equation*}
  with $\mathbf{Q}^{(k)} = (\mathbf{X}^{(h)T}\mathbf{\Omega}^{(k)}\mathbf{X}^{(k)} + \mathbf{Q}^{-1})^{-1} $ and
  $\mathbf{b}^{(k)} = \mathbf{Q}^{(k)}(\mathbf{X}^{(k)T} \boldsymbol{\kappa}^{(k)} + \mathbf{Q}^{-1} \mathbf{b})$.
  }
  \ENDFOR
  \STATE -----------------------------------------------------------------------------------------------------
  \STATE { [4] Allocate each birth defect $i$ to one of the clusters}
  \FOR{$i = 1, \ldots, N$}
  \STATE Sample the class indicator $c_i$ conditionally on $\boldsymbol{c}_{-i} = (c_1, \ldots, c_{i - 1}, c_{i +1 }, \ldots, c_n)$ from the discrete distribution with probabilities
  \begin{align*}
   \Pr(c_i= k| \boldsymbol{c}_{-i}, - ) \propto \Pr(c_i= k| \boldsymbol{c}_{-i}) \Pr(\mathbf{y}_{i}|\mathbf{X}_i,\alpha_i, c_i = k, \boldsymbol{\beta}^{(k)})
  \end{align*}
  with
  \begin{equation*}
   \Pr(\mathbf{y}_{i}| \mathbf{X}_i, \alpha_i, c_j = k, \boldsymbol{\beta}^{(k)}) = \prod_{j = 1}^{n_i} \left[ \exp(\alpha_i + \mathbf{x}_{ij}^{T} \boldsymbol{\beta}^{(k)})^{y_{ij}} \right] \left[ 1 +  \exp(\alpha_i + \mathbf{x}_{ij}^{T} \boldsymbol{\beta}^{(k)}) \right]^{(-1)}
  \end{equation*}
  being the model likelihood evaluated for cluster $h$ and $\Pr(c_i= h| \boldsymbol{c} ^{ -(i)})$ computed as described in Section~\ref{sec4:Posterior computation}.
  \ENDFOR
 \end{algorithmic}
 \label{Algorithm_1}
\end{breakablealgorithm}
\newpage


\subsection*{Results for NBDPS data application}

\begin{figure}[H]
 \includegraphics[scale = 0.40]{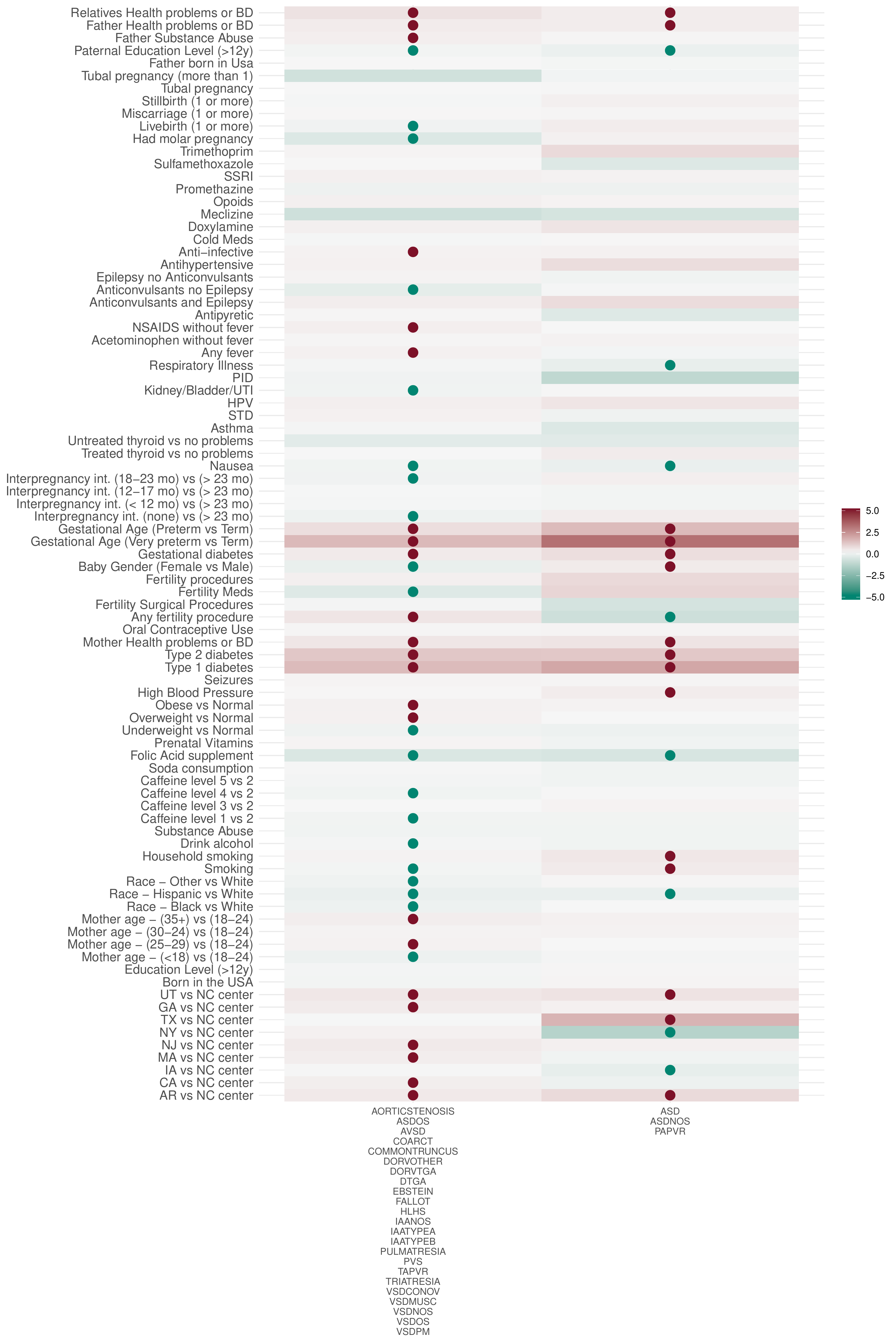}
 \caption{\textbf{CP process with $\psi = 0$}. Posterior mean estimates of log odds-ratios, with values shown if significant at $95\%$ using credibility intervals. Labels on the x-axis list the defects in each cluster. Red color indicates a risk factor with green a protective effect.}
 \label{fig:odds_dp}
\end{figure}

\begin{figure}[H]
 \includegraphics[scale = 0.40]{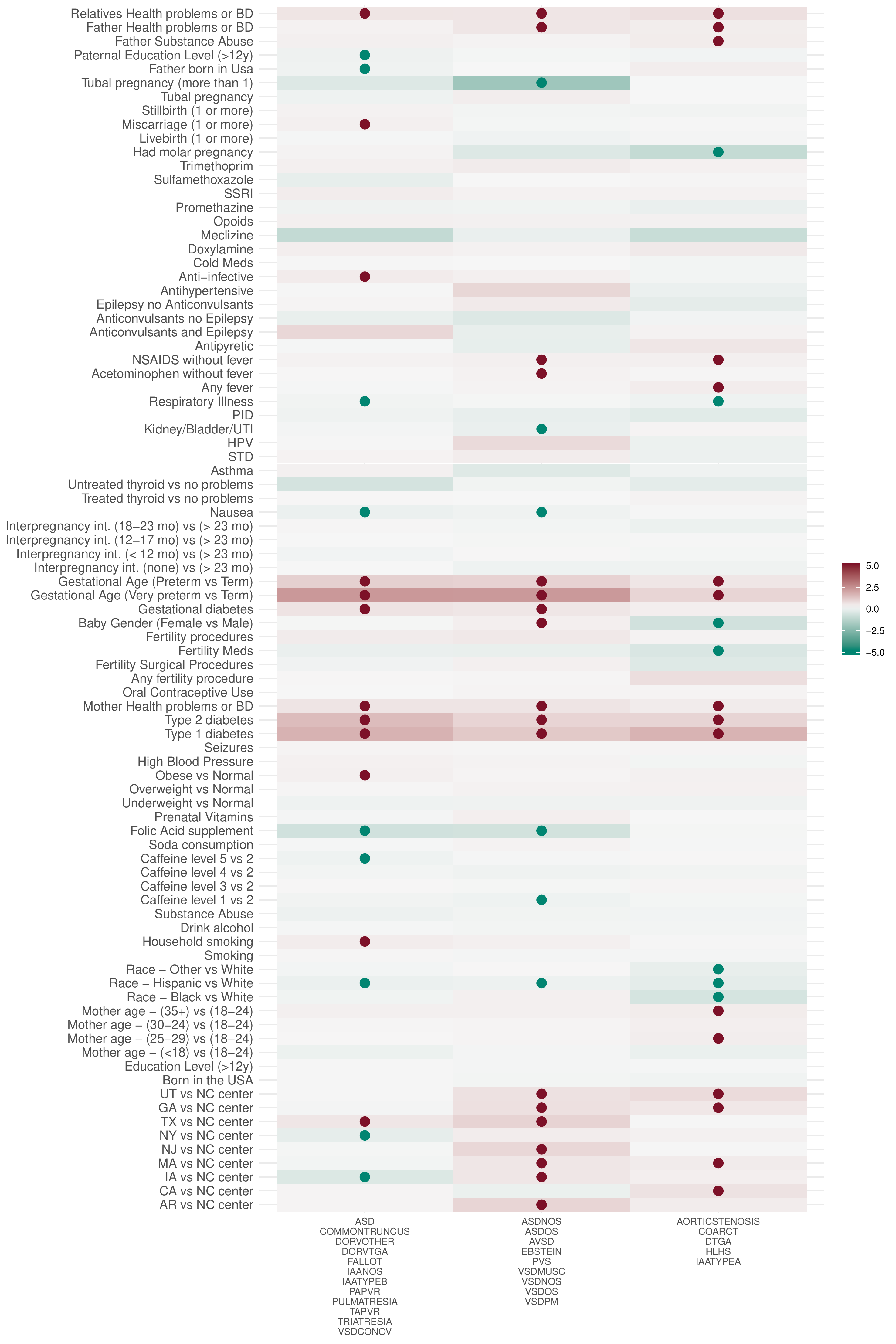}
 \caption{\textbf{CP process with $\psi = 40$}. Posterior mean estimates of log odds-ratios, with values shown if significant at $95\%$ using credibility intervals. Labels on the x-axis list the defects in each cluster. Red color indicates a risk factor with green a protective effect. }
 \label{fig:cp_1}
\end{figure}

\begin{figure}[H]
 \includegraphics[scale = 0.40]{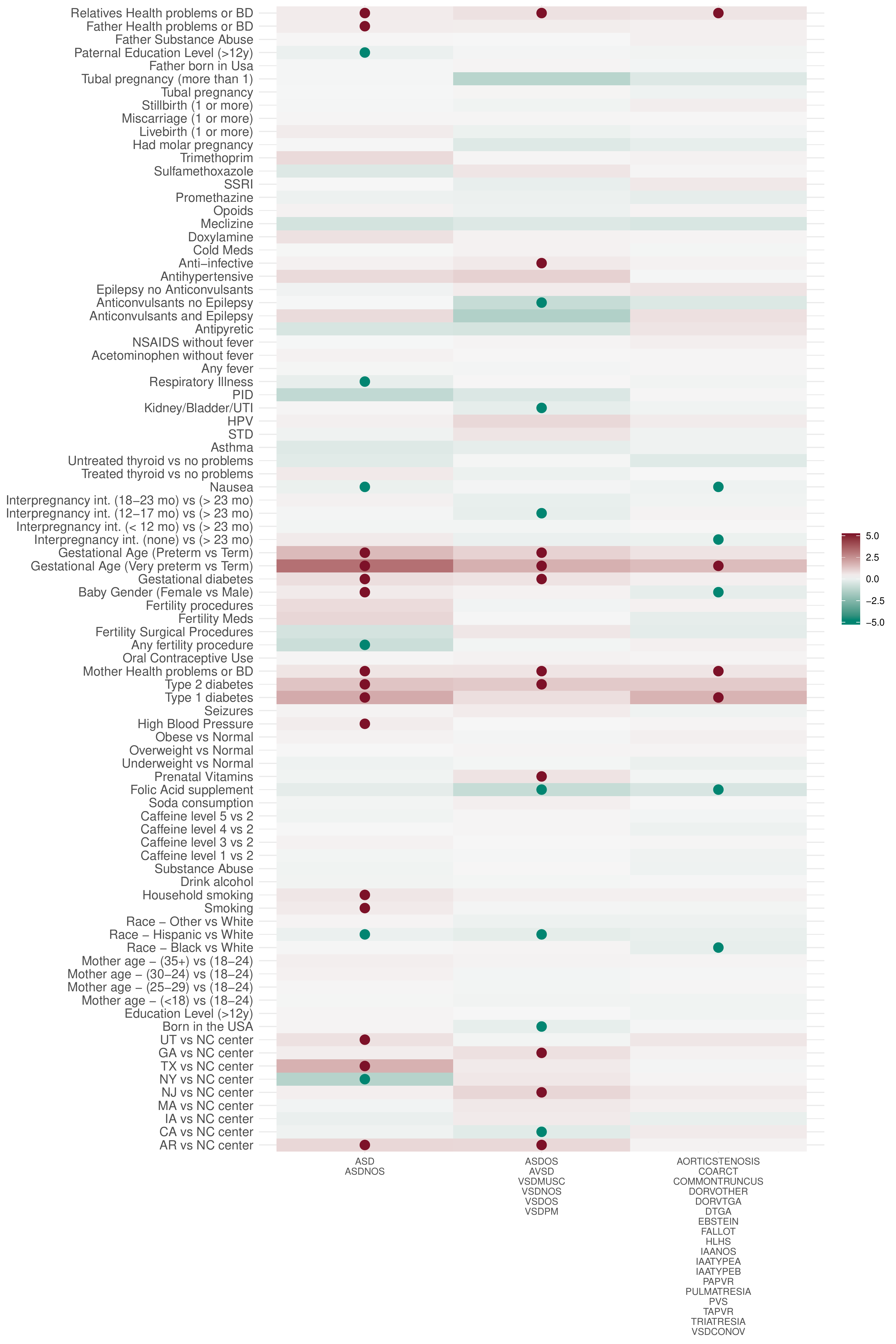}
 \caption{\textbf{CP process with $\psi = 80$}. Posterior mean estimates of log odds-ratios, with values shown if significant at $95\%$ using credibility intervals. Labels on the x-axis list the defects in each cluster. Red color indicates a risk factor with green a protective effect. }
 \label{fig:cp_2}
\end{figure}

\begin{figure}[H]
 \includegraphics[scale = 0.40]{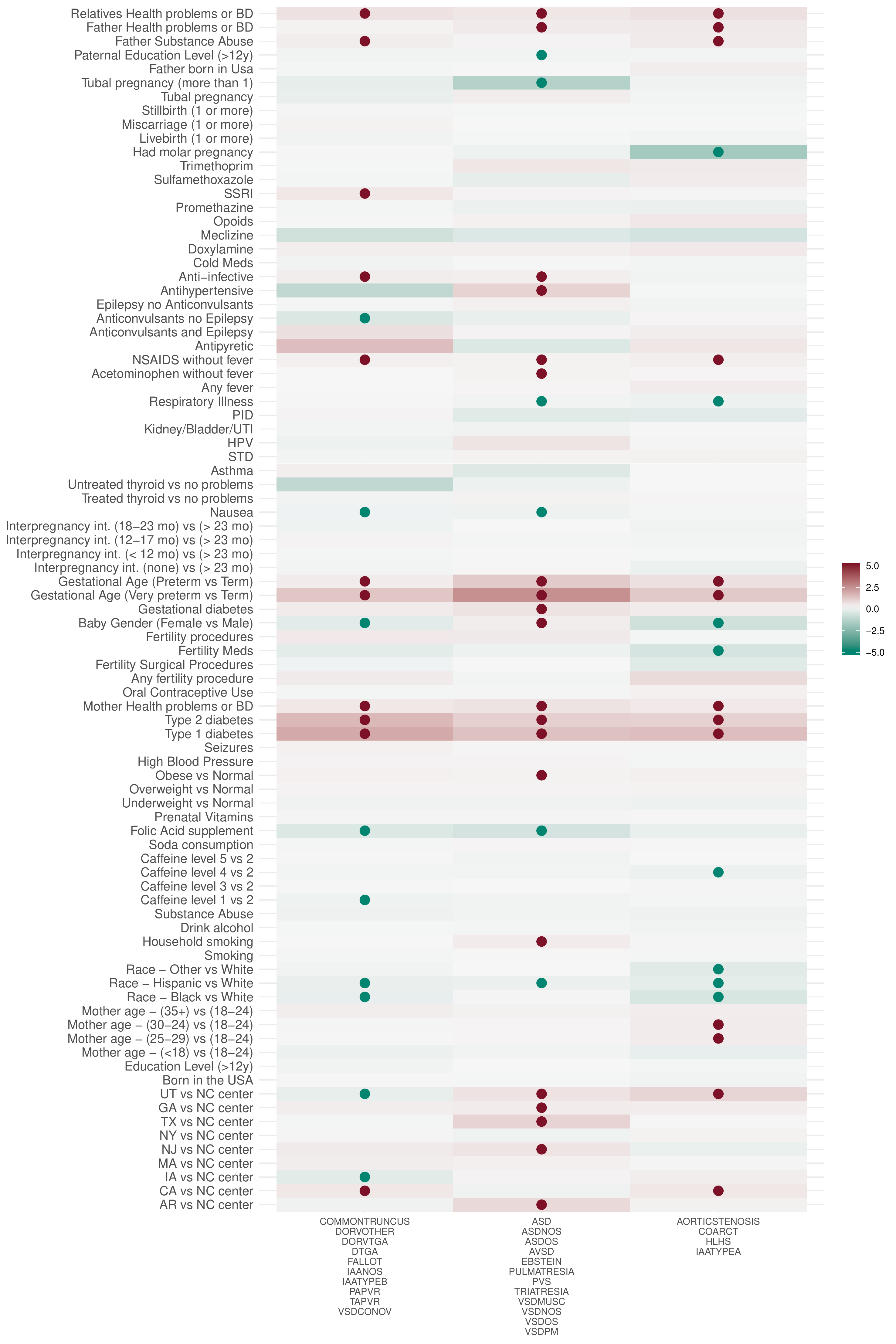}
 \caption{\textbf{CP process with $\psi = 120$}. Posterior mean estimates of log odds-ratios, with values shown if significant at $95\%$ using credibility intervals. Labels on the x-axis list the defects in each cluster. Red color indicates a risk factor with green a protective effect.}
 \label{fig:cp_3}
\end{figure}

\begin{figure}[H]
 \includegraphics[scale = 0.40]{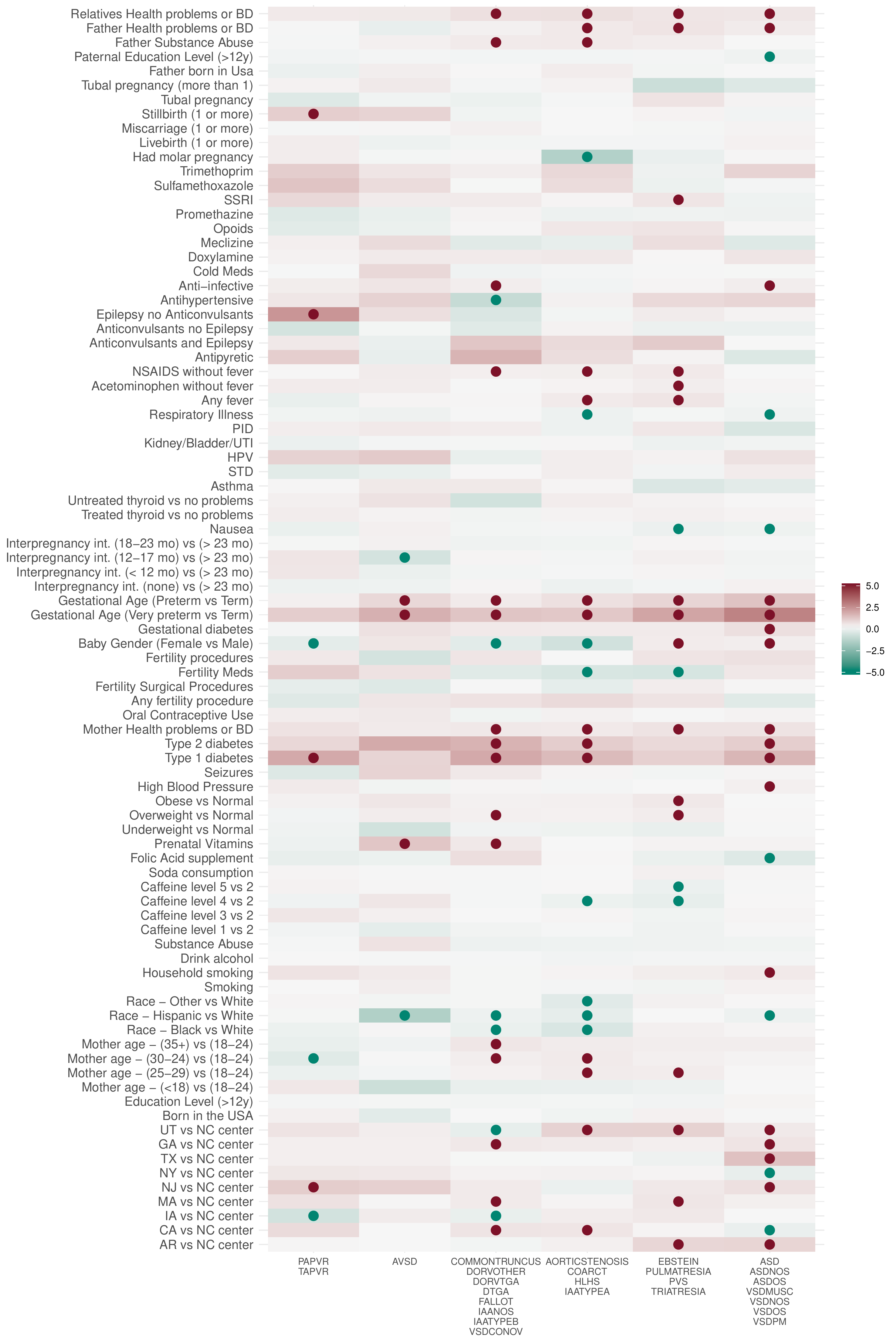}
 \caption{\textbf{CP process with $\psi = \infty$}. Posterior mean estimates of log odds-ratios, with values shown if significant at $95\%$ using credibility intervals. Labels on the x-axis list the defects in each cluster. Red color indicates a risk factor with green a protective effect.}
 \label{fig:pg}
\end{figure}

\end{document}